\documentclass[aps,prl,twocolumn,superscriptaddress,10pt,showpacs]{revtex4-1}
\usepackage{graphicx}
\usepackage{amsfonts}
\usepackage{amsmath}
\usepackage{amssymb}
\usepackage{epsfig}
\usepackage{dsfont}
\usepackage{color}
\newcommand{\kt}[1]{\ensuremath{\left|#1\right\rangle}}
\newcommand{\br}[1]{\ensuremath{\left\langle#1\right|}}

\newcommand{\ket}[1]{| #1\rangle}

\newcommand{\deltacav}{\delta_c}
\newcommand{\deltaryd}{\delta_R}
\newcommand{\gammaryd}{\gamma_R}
\newcommand{\thetadark}{\theta}
\newcommand{\gammadark}{\gamma_D}
\newcommand{\gammastark}{\gamma_b}
\newcommand{\deltadark}{\delta_D}

\newcommand{\Tmax}{T_{\mathrm{max}}}

\newcommand{\pol}{\alpha_n}

\newcommand{\SIfull}{Supplemental Material}

\newcommand{\uchicago}{Department of Physics and James Franck Institute, University of Chicago, Chicago, IL}

\begin{document}
\title{Observation of Cavity Rydberg Polaritons}
\author{Jia Ningyuan}
\author{Alexandros Georgakopoulos}
\author{Albert Ryou}
\author{Nathan Schine}
\author{Ariel Sommer}
\author{Jonathan Simon}
\affiliation{\uchicago}
\date{\today}
\begin{abstract}
We demonstrate hybridization of optical cavity photons with atomic Rydberg excitations using electromagnetically induced transparency (EIT). 
The resulting dark state Rydberg polaritons exhibit a compressed frequency spectrum and enhanced lifetime indicating strong light-matter mixing. 
We study the coherence properties of cavity Rydberg polaritons and identify the generalized EIT linewidth for optical cavities.
Strong collective coupling suppresses polariton losses due to inhomogeneous broadening, which we demonstrate by 
using different Rydberg levels with a range of polarizabilities.
Our results point the way towards using cavity Rydberg polaritons as a platform for creating photonic quantum materials.
\end{abstract}

\maketitle

Coupling photons to electronic excitations of a medium leads to hybrid quasiparticles, or polaritons, that carry properties of both light and matter. The photonic component allows polaritons to propagate like light, while the material component enables interactions between polaritons. An important example is exciton polaritons in semiconductor microcavities, which exhibit an effective mass and two-dimensional motion arising from the photonic component, while the exciton component leads to a mean-field interaction, allowing Bose-Einstein condensation~\cite{bali2007bose,deng2010exci,caru2013quan}.
Rydberg polaritons in atomic gases enable strong interactions even at the few-quantum level~\cite{luki2001dipo,prit2010coop,gors2011phot,dudi2012stro,peyr2012quan, maxw2013stor,hofm2013sub-, firs2013attr, bien2014scat}, a key ingredient for producing highly correlated states, including fractional quantum Hall states~\cite{grus2013frac,hafe2013non-,umuc2014prob,magh2015frac,somm2015quan} and emergent quantum crystals~\cite{somm2015quan,chan2008crys,gopa2009emer,otte2013wign,koll2015adju}. 
While previous work on Rydberg polaritons has focused on one-dimensional free-space light fields, photons in optical cavities provide access to two-dimensional motion, harmonic trapping~\cite{klae2010bose}, and effective magnetic fields~\cite{somm2015quan, somm2015engi}. In addition, optical cavities can enhance the optical non-linearity arising from Rydberg interactions~\cite{pari2012obse,stan2013disp}.

Rydberg polaritons are formed by coherently coupling light to a highly excited atomic Rydberg level using electromagnetically induced transparency (EIT)~\cite{luki2003coll}. 
At EIT resonance, destructive interference prevents population of a lossy intermediate atomic level, resulting in a 
dark state polariton~\cite{flei2000dark} that consists of a 
superposition of a photon and a collective atomic Rydberg excitation. 
A large admixture of the (long-lived) atomic excitation in a dark state polariton slows all photonic dynamics~\cite{flei2000dark}.
In an optical cavity, this results in a polariton whose lifetime 
can exceed the empty-cavity lifetime by orders of magnitude, and an energy that is pulled toward the EIT resonance~\cite{luki1998intr, wang2000cavi,hern2007vacu,wu2008obse,mu2010elec,kamp2014elec}. 
In a multimode cavity, hybridization rescales the trap frequency and effective mass of the polariton~\cite{somm2015quan}.

\begin{figure}[b]
\includegraphics{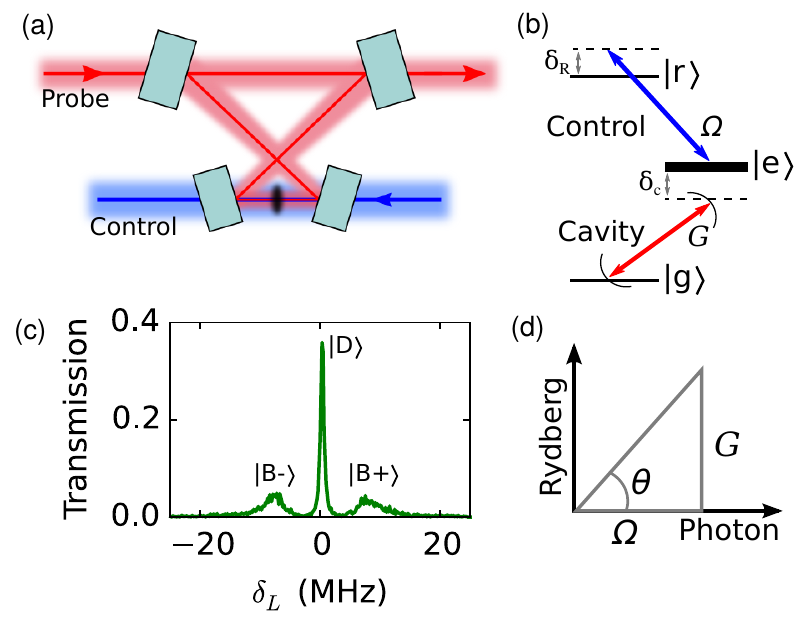}
\caption{(Color online) Cavity Rydberg EIT. (a) Schematic of the experimental setup, showing the $^{87}$Rb atomic sample in the waist of a running-wave cavity. The cavity mirrors have a high finesse at the probe laser wavelength (780 nm) while being anti-reflection coated for the control laser (480 nm) that counter-propagates through the sample. (b) Level diagram showing that the cavity mode couples the atomic ground state to an excited state, with collective vacuum Rabi frequency $G$ and detuning $\deltacav$, while the control laser couples the excited state to a Rydberg level, with Rabi frequency $\Omega$ and detuning $\deltaryd$. (c) Transmission spectrum measured at $\deltacav=\deltaryd=0$ showing peaks due to the Rydberg polariton dark state $\kt{D}$ and the two bright polariton states $\kt{B\pm}$. Here $G$=13 MHz, $\Omega$=7 MHz. (d) Composition of the dark polariton, a superposition of a cavity photon and a collective Rydberg excitation, given by the couplings $G$ and $\Omega$ that define the dark state rotation angle $\thetadark$.
}
\label{fig:setup}
\end{figure}

\begin{figure*}[t]
\includegraphics{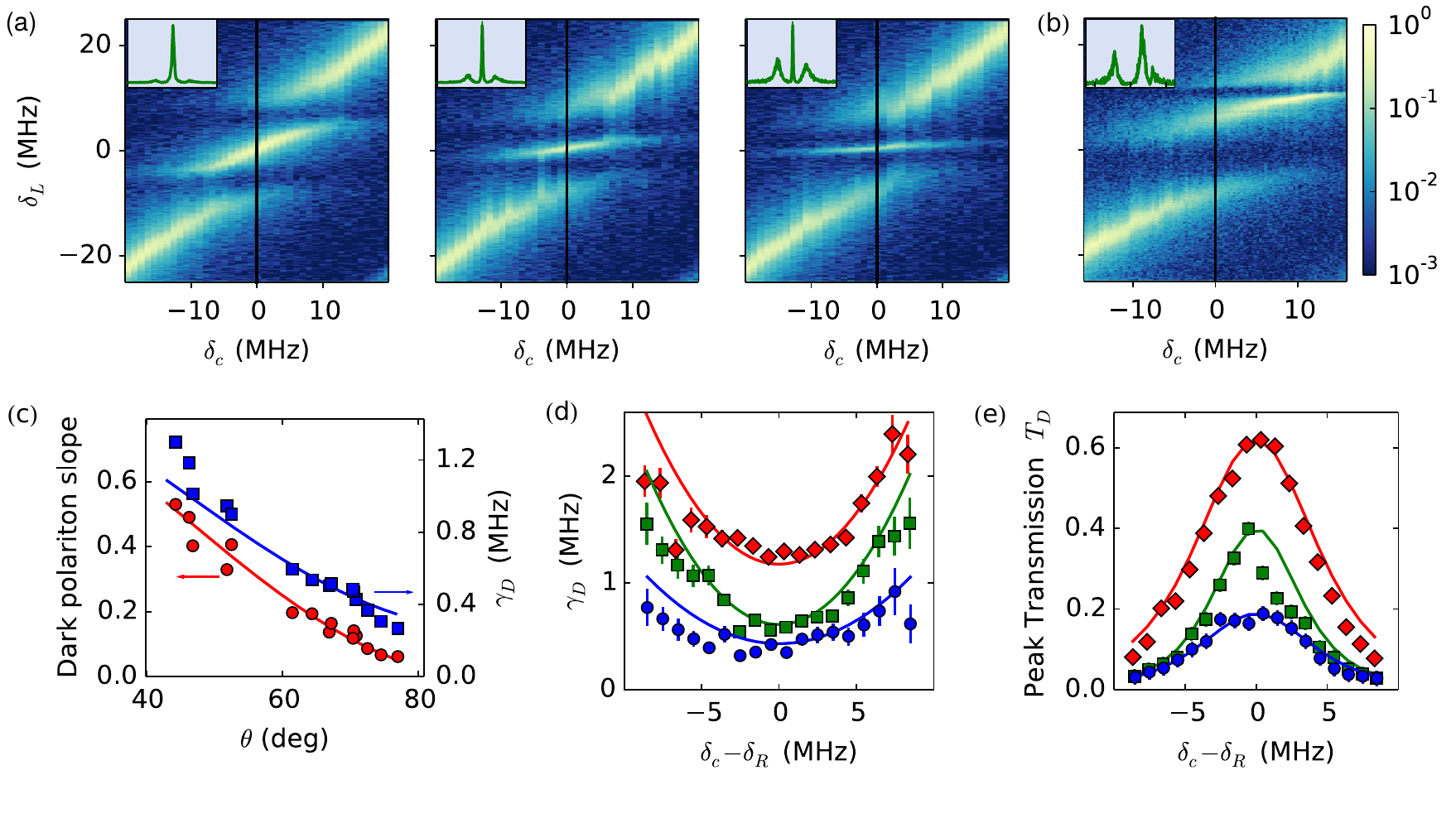}
\caption{(Color online) Spectroscopy of cavity Rydberg polaritons. (a) Cavity transmission spectra as a function of cavity detuning $\deltacav$ for several values of the control laser power. From left to right, 
$\thetadark$ (deg)=43, 62, 72; 
$\Omega$ (MHz)= 13.1(1), 6.9(1), 4.9(1); 
$G$ (MHz) = 12.3(2), 13.0(1), 14.7(1).
Here $|\delta_R|<1$ MHz. Insets in (a) and (b): spectra along the vertical line at $\deltacav=0$. Color scale for (a) and (b): cavity transmission as a fraction of the empty-cavity peak transmission.
(b) Transmission spectrum at non-zero control laser detuning $\delta_R=9.8(4)$ MHz. Here  $\Omega$ (MHz) = 8.2(6), $G$ (MHz) = 16.8(3). 
(c) Energy and lifetime versus dark state rotation angle. Red circles (left): dark polariton slope $d\deltadark/d\deltacav$. Blue squares (right): dark polariton inverse lifetime $\gammadark$.  Solid lines: theoretical predictions, using $\Omega$ obtained from the calibrated control laser power and $G$ obtained from fitting to the transmission spectrum. (d) and (e) show the effect of detuning the cavity from EIT resonance using the data in (a). Correspondence to (a): left--diamonds (red), middle--squares (green), right--circles (blue).
(d) $\gammadark$ versus cavity detuning from EIT resonance. Solid lines: second-order prediction (\ref{eq:darkgamma}), using $\gammaryd$ and $G$ obtained from the transmission spectrum at $\deltaryd-\deltacav$=0. (e) 
Height $T_D$ of the dark polariton peak versus cavity detuning. Solid lines: theoretical prediction using (\ref{eqn:EITwidth}) plus higher order corrections~\cite{si} that are only significant for the lower (blue) curve.
}
\label{fig:bigspectra}
\end{figure*}

We experimentally observe Rydberg polaritons in an optical cavity and 
explore the spectral and coherence properties of these collective states in cavity transmission spectroscopy.
While Doppler decoherence and inhomogeneous Stark shifts~\cite{taus2010spat,hatt2012detr,hoga2012driv} are more significant for Rydberg levels than in $\Lambda$-type configurations previously considered~\cite{wu2008obse}, we observe that
collective coupling suppresses decoherence arising from inhomogeneous broadening and we provide a theoretical interpretation of this effect.
Finally, we demonstrate Rydberg EIT simultaneously in two nearly-degenerate cavity modes, showing that frequency pulling~\cite{luki1998intr} leads to a compression of the mode spectrum, and pointing the way to many-mode, many-body experiments~\cite{somm2015quan}.

Our system consists of an ensemble of ground-state atoms in the waist of an optical cavity, as illustrated in Fig.~\ref{fig:setup}(a). 
A cavity photon ($\kt{c}$) may be absorbed by the atomic ensemble, generating a collective excitation $\kt{e_c}$ of the $\kt{e}$ level, 
with collective vacuum Rabi frequency $G$. 
A control laser transfers the $\kt{e_c}$ state to a collective excitation $\kt{r_c}$ of the Rydberg level $\kt{r}$ with Rabi frequency $\Omega$. 
The eigenstates of the coupled atom-cavity system include a dark state $\kt{D}$ and two bright states $\kt{B\pm}$. When the cavity is tuned to the EIT resonance condition $\deltacav=\deltaryd$, with $\deltacav$ the detuning of the cavity from the atomic transition and $-\deltaryd$ the detuning of the control laser from $\kt{e}$ to the Rydberg level, the dark state is 
$\kt{D}=\cos\thetadark\kt{c}-\sin\thetadark\kt{r_c}$, where $\thetadark\equiv\tan^{-1}(G/\Omega)$ is the dark state rotation angle~\cite{flei2005elec}. For small detuning from EIT resonance, and including losses, the energy $\hbar\deltadark$ and inverse lifetime $\gammadark$ of the dark polariton are given by
\begin{eqnarray}
	\deltadark &\approx &  \deltacav\cos^2\thetadark + \deltaryd \sin^2 \thetadark\label{eq:darkfreq}\\ 
	\gammadark &\approx & \kappa\cos^2\thetadark + \gammaryd \sin^2\thetadark +a\left(\deltacav-\deltaryd\right)^2\label{eq:darkgamma}
\end{eqnarray}
with $a={4\Omega^2 G^2\Gamma}/{(\Omega^2+G^2)^3}$, and $\delta_D$ measured relative to the atomic transition frequency. Here $\kappa$, $\Gamma$, and $\gamma_R$ are the linewidths of $\kt{c}$, $\kt{e_c}$, and $\kt{r_c}$, respectively.

To observe Rydberg polaritons in an optical cavity, we prepare
a laser-cooled $^{87}$Rb atomic sample and transport it into the waist of a running-wave bow-tie optical cavity.
The cavity is tuned near the atomic D$_2$ transition (780 nm) from the $\kt{g}$ = 5S$_{1/2}$(F=2) ground state to the $\kt{e}$ = 5P$_{3/2}$(F=3) excited state. The control laser (wavelength 480 nm, waist 29 $\mu$m) that couples to the $\kt{r}$ = $nS_{1/2}$ Rydberg level counter-propagates through the sample. Here we use primarily $n=40$.  
At 780 nm, the cavity has a TEM$_{00}$ mode waist of 12 by 11 $\mu$m ($1/e^2$ intensity radii) between the lower mirrors, and a finesse of 2500 (1.8 MHz FWHM linewidth). We obtain transmission spectra by sweeping the detuning $\delta_L$ (relative to the atomic resonance) of a probe laser linearly over 1 ms and detecting the transmitted light with a single-photon counter (Fig. \ref{fig:setup}c). Fitting the theoretical lineshape~\cite{si} 
to the spectrum provides an estimate of $G$, $\Omega$, and $\gammaryd$. 
We use the $\Omega$ determined from fitting to obtain a global calibration $\Omega=b \sqrt{P_c}$, where $P_c$ is the control laser power~\cite{si}.

\begin{figure}
\includegraphics{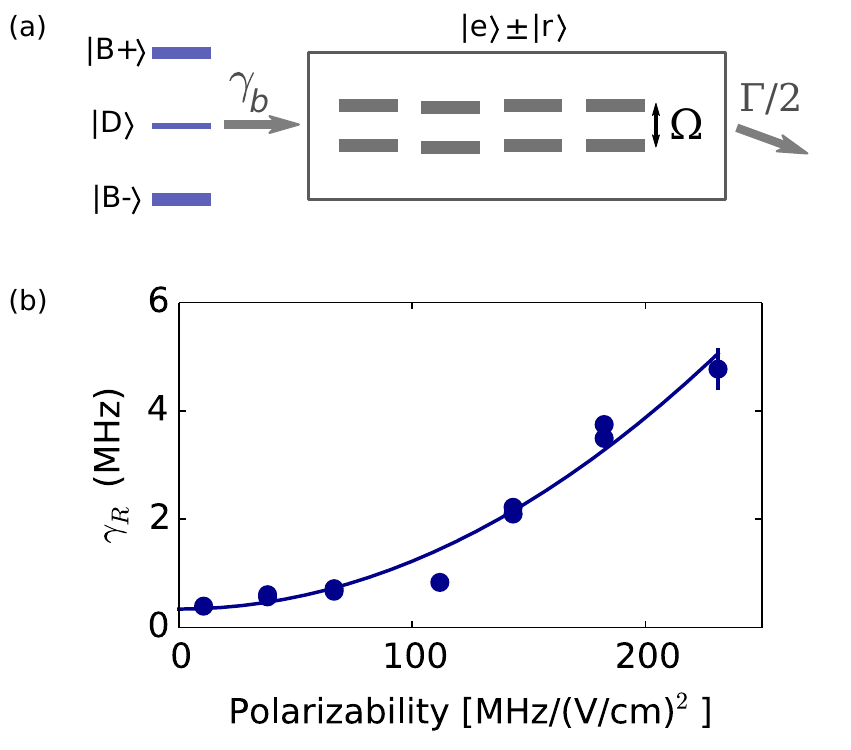}
\caption{(Color online) Collective suppression of decoherence. (a) Inhomogeneous broadening couples the dark polariton to a bath of states orthogonal to the cavity mode. The splitting from the control laser detunes these states from the dark state, suppressing the lossy channel. (b) Effective decay rate $\gammaryd$ of the collective Rydberg state for varying principle quantum number $n$ versus the polarizability of the Rydberg state. Here $n$=40, 48, 52, 56, 58, 60, and 62. $\Omega$ (MHz)=7.8(5), $G$ (MHz) = 12(2), average and std. dev. over the data sets. Solid line: quadratic fit.}
\label{fig:decoherence}
\end{figure}

In Fig. \ref{fig:bigspectra}, we observe the energy and inverse lifetime of the dark polariton versus cavity frequency across the EIT resonance. 
The energy of the dark polariton is given by the probe detuning $\deltadark$ at the narrow peak.
As the cavity detuning is varied in Fig. \ref{fig:bigspectra}(a), $\deltadark$ varies with a slope $d\deltadark/d\deltacav < 1$, demonstrating frequency-pulling toward EIT resonance~\cite{luki1998intr,hern2007vacu}.
Figure \ref{fig:bigspectra}(c) shows the slope $d\deltadark/d\deltacav$ near $\deltacav=0$, versus the dark state rotation angle, demonstrating good agreement with the predicted slope of $\cos^2\thetadark$ given by (\ref{eq:darkfreq}).
The right axis in Fig. \ref{fig:bigspectra}(c) shows the inverse lifetime $\gammadark$ obtained from the full-width at half-maximum of a Lorentzian fit to the dark polariton peak with the cavity at EIT resonance. 
Here $\gammadark$ reaches a minimum of 0.26(1) MHz, significantly less than the cavity linewidth.

Tuning the cavity away from the EIT condition $\deltacav=\deltaryd$
mixes the lossy state $\kt{e_c}$ into the dark polariton.
Figure \ref{fig:bigspectra}(d) shows a quadratic dependence of $\gammadark$ on $\deltacav-\deltaryd$ that agrees well with (\ref{eq:darkgamma}).  
To further explore the concept of EIT linewidth in an optical cavity, we look in Fig.~\ref{fig:bigspectra}(e) at the height $T_D$ of the dark polariton peak, versus the detuning of the cavity from EIT resonance.
The data follow a squared Lorentzian~\cite{si}
$T_D \propto \left[  1+(\deltacav-\deltaryd)^2/(\Gamma_w/2)^2  \right]^{-2}$.
The width $\Gamma_w$ provides a notion of the EIT linewidth in an optical cavity.  To lowest approximation,
\begin{equation}
	\Gamma_w = \frac{\Omega^2+G^2}{G}\sqrt{\frac{\kappa}{\Gamma}}
	\label{eqn:EITwidth}
\end{equation}
for small $\gammaryd$ and $\kappa$. 
The relation of $\Gamma_w$ to the free-space EIT linewidth is discussed in the \SIfull~\cite{si}.

Compared to ground-state cavity EIT, Rydberg polaritons have a greater sensitivity to Doppler broadening (due to the wavelength mismatch in the ladder configuration) and inhomogeneous electric field shifts. 
To isolate the dephasing due to a stray electric field gradient, we vary the polarizability $\pol$ of the  Rydberg level by changing the principle quantum number $n$. Figure \ref{fig:decoherence}(b) shows the loss rate $\gammaryd$ of $\kt{r_c}$ versus $\pol$, obtained by fitting the measured spectra.
For spectroscopy of completely independent atoms, one would expect $\gammaryd$ to vary like the inhomogeneous broadening $\gammastark\propto \alpha_n$ of the Rydberg level.
However, the data in Fig.~\ref{fig:decoherence}(b) vary quadratically with $\alpha_n$ due to a collective enhancement of coherence. 
Dephasing couples $\kt{r_c}$ to a bath of collective excitations~\cite{hone2011arti} of the coupled $\kt{r}$ and $\kt{e}$ levels, that, in turn, have no coupling to the cavity mode. In the rotating frame, these states are detuned by $\pm \Omega/2$ relative to the dark polariton (Fig. \ref{fig:decoherence}a). For $\Omega > \Gamma$, this leads to a suppressed loss rate of 
$s_b\gammastark^2 \Gamma/\Omega^2$ from $\kt{r_c}$, for some constant $s_b$. From numerical simulations~\cite{si}, we find $s_b\approx 4$, for a normal distribution of Stark shifts with standard deviation $\gammastark$. The simulations confirm  that Doppler decoherence is suppressed by the same mechanism.

The finite waist size of the control laser leads to inhomogeneous couplings between the excited state and the Rydberg level.  
However, the inhomogeneity simply modifies $\kt{r_c}$; a dark state still exists~\cite{si}. The (nearly un-populated) $\kt{e_c}$ state now couples to the bath of orthogonal states, effectively increasing $\Gamma$. Numerical simulations confirm that the inhomogeneity of the control laser intensity does not cause significant dephasing of the dark state~\cite{si}.


\begin{figure}[t]
\includegraphics{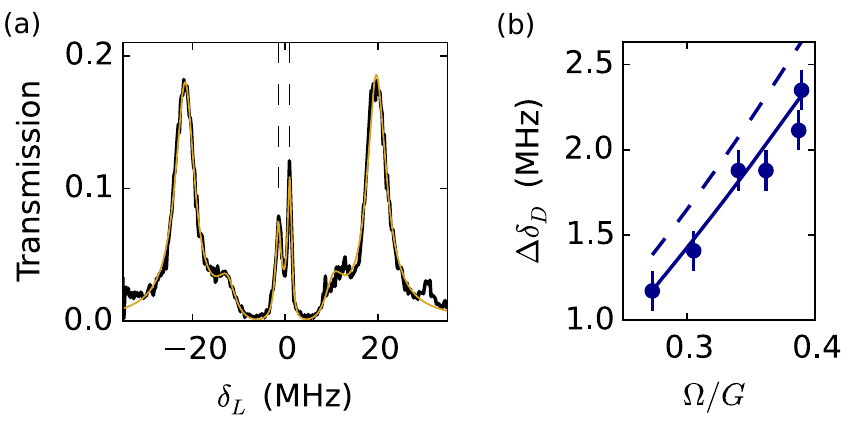}
\caption{(Color online) Rydberg polaritons in two cavity modes. The TEM$_{02}$ and TEM$_{10}$ modes are tuned to $\pm$10 MHz from EIT resonance. (a) Transmission spectrum for $\Omega/G$=0.39. The dark  polariton resonances (dashed lines) are separated by $\Delta\deltadark=$ 2.3 MHz. Curve: fit to theoretical model for two orthogonal collective states. (b) 
Adjusting the control laser power varies the photonic component of the dark polariton states, tuning $\Delta\delta_D$. Dashed curve: first-order prediction (\ref{eq:darkfreq}). Solid curve: numerical solution. }
\label{fig:twomode}
\end{figure}

We demonstrate the implications of our results for multimode experiments by implementing Rydberg EIT simultaneously in two cavity modes. 
The cavity has TEM$_{02}$ and TEM$_{10}$ modes nearly degenerate, being separated by 20 MHz~\cite{si}. Figure \ref{fig:twomode}(a) shows a measured transmission spectrum for two-mode cavity Rydberg EIT with the TEM$_{02}$ and TEM$_{10}$ modes tuned to either side of the EIT resonance. The two central peaks correspond to dark polaritons in the two modes and are separated by much less than 20 MHz, due to strong light-matter mixing.
Figure \ref{fig:twomode}(b) shows that the splitting between the two dark polariton peaks can be tuned by varying the laser couplings.  

Our results imply an intuitive picture for dark polaritons in multimode cavities. For a cavity with mode frequencies $\deltacav^{(n)}$ relative to EIT resonance, each mode within the cavity EIT linewidth $\Gamma_w$ supports a long-lived dark polariton of energy $\hbar\deltadark^{(n)}=\hbar\deltacav^{(n)}\cos^2\thetadark$. As the dynamics of cavity photons in the transverse plane are given by an effective Hamiltonian $H_{\mathrm{ph}}=\sum_n\kt{n}\hbar\deltacav^{(n)}\br{n}$, the effective Hamiltonian for transverse dynamics of a dark polariton is then $H_{\mathrm{ph}}\cos^2\thetadark$~\cite{somm2015quan}. The number of modes within $\Gamma_w$ determines the spatial resolution and number of degrees of freedom.

In conclusion, we have observed Rydberg polaritons in an optical cavity, studied their energy and coherence properties, and extended the concept of EIT linewidth to cavity EIT. In addition, we demonstrated a novel mechanism for collective suppression of decoherence and observed cavity Rydberg EIT in a two-mode optical cavity. 
Introducing polariton-polariton interactions by working with higher Rydberg levels provides a route toward next-generation single photon sources~\cite{dudi2012stro,peyr2012quan} and single photon transistors~\cite{gorn2014sing,baur2014sing, tiar2014sing} in an optical cavity. In addition, nearly-degenerate multimode cavities can provide a spatial degree of freedom for implementing condensed-matter models using Rydberg polaritons, where the increasing loss rate for large mode detunings provides a potential mechanism for evaporative cooling into many-body states.

\begin{acknowledgments}
The authors thank Lindsay Bassman, Graham Greve, Aaron Krahn, Sohini Upadhyay, Jin Woo Sung, Michael Cervia, and William Tahoe Schrader for contributions to the experimental system. The U.S. AFOSR (grant FP053419-01-PR) supported apparatus construction and the U.S. DOE (grant FP054241-01-PR) supported theoretical modeling.
\end{acknowledgments}
%
%

\clearpage
\onecolumngrid
\section{Supplemental Material}

\begin{figure}
  \includegraphics[width=140mm]{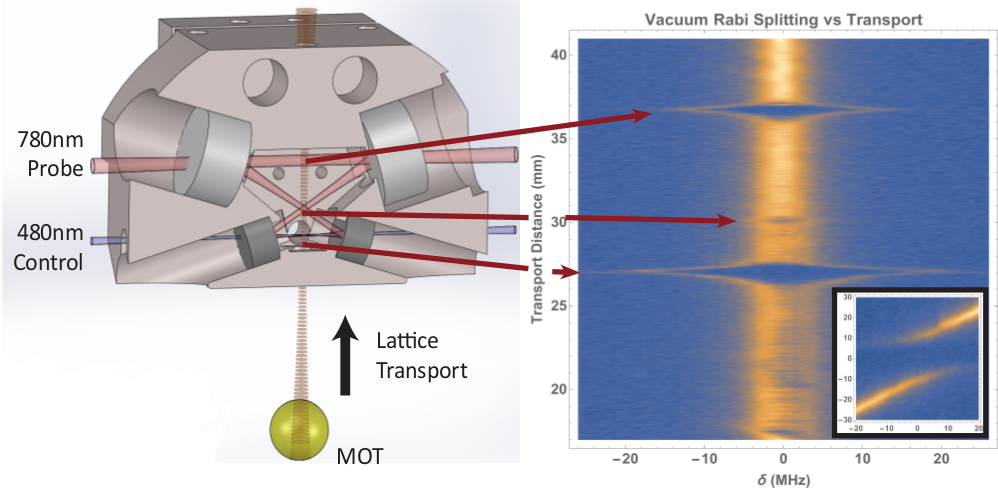}
  \caption{\textbf{Cavity geometry, and vacuum Rabi splitting vs. atom transport distance}. (\textbf{Left}) The four-mirror optical resonator has two nearly flat (500 mm ROC) upper mirrors, and two 10 mm ROC lower mirrors. The cavity is locked (at 1560 nm), and the probe beam (at 780 nm) is injected and detected through the upper mirrors, which exhibit larger transmission at 780 and 1560 nm. The 480 nm Rydberg coupling laser is injected through the lower mirrors, which are AR-coated at 480 nm. (\textbf{Right}) As the atomic cloud is transported through the optical resonator, vacuum Rabi splittings appear at each crossing of the atoms with the TEM$_{00}$ resonator mode. The lower splitting is the atomic crossing with the small (12 by 11 $\mu$m) waist, the upper splitting is the crossing with the large (94 by 58 $\mu$m) waist. The small splitting in the middle arises when the atomic cloud passes \emph{between} the two intermediate waists- the space between these waists arises from a small twist (non-planarity) of the resonator.}
  \label{fig:CavitySetupVRS}
\end{figure}

\begin{figure}
\includegraphics[scale=0.9]{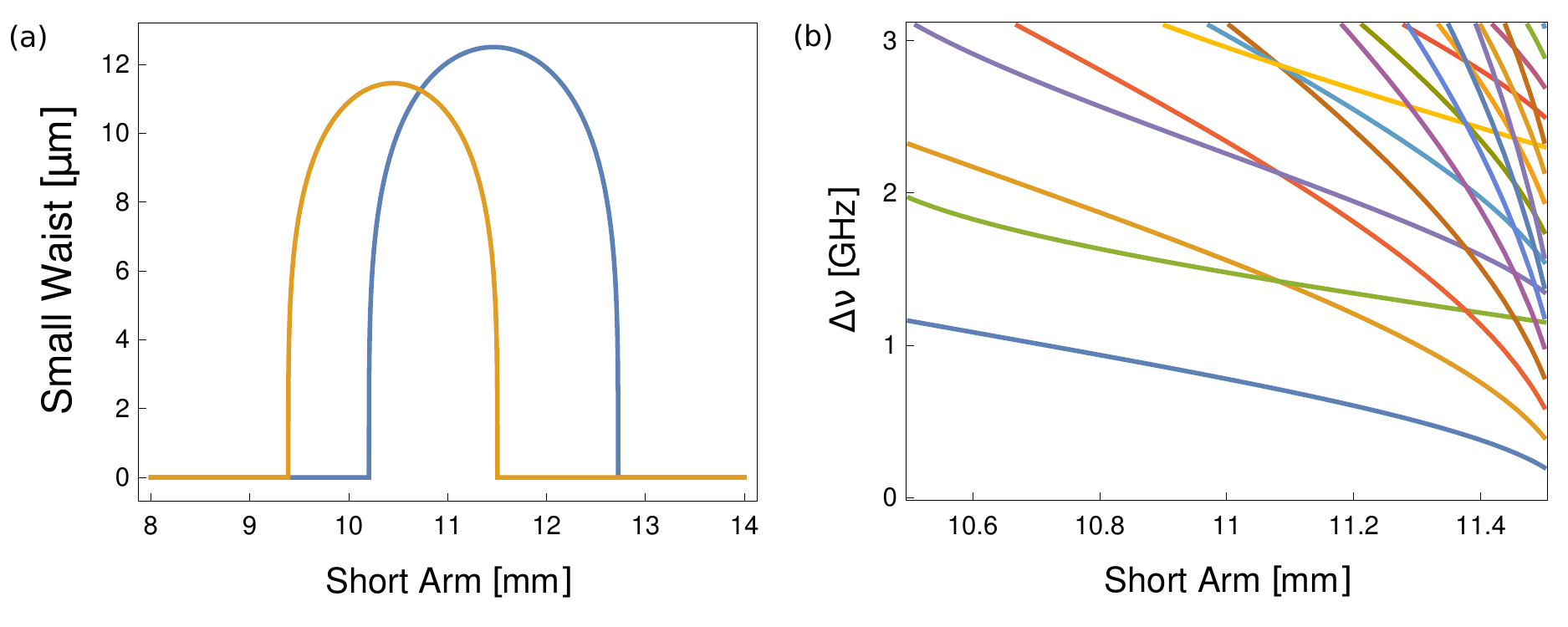}
  \caption{\textbf{Resonator stability and transverse mode degeneracy.} (a) Waist sizes in sagittal (orange) and tangential (blue) planes between the 10 cm-ROC mirrors (short arm) as a function of the length of the short arm. Unlike a typical two-mirror resonator, where the stability region extends from a length of zero (the ``planar limit'') to a length of twice the mirror radii of curvature (the ``concentric limit''), a four-mirror Ti:Saph-style resonator has a substantially reduced stability region around a short-arm length given approximately by the radii of curvature of the highly curved mirrors; in exchange for this reduced stability, one obtains a smaller mode waist with \emph{reduced} stability requirements compared with acquiring this mode waist in a near-concentric resonator. We operate near a length of 11 mm, where both saggital and tangential axes of the mode are relatively stable; the shift between the axes arises from astigmatism induced by non-normal incidence on the curved resonator mirrors \cite{sieg1986lase}. (b) Transverse mode frequencies relative to the TEM$_{00}$ mode. The curve closest to the x-axis (blue) corresponds to TEM$_{01}$.  The next two lowest curves, green and orange, correspond to TEM$_{02}$ and TEM$_{10}$, respectively.  Near a short arm length of 11.1 mm is the degeneracy point at which $\nu_{n,m+2,q} = \nu_{n+1,m,q}$.  }
  \label{fig:CavityStability}
\end{figure}

\section*{Experimental Details}
\subsection*{Cavity Geometry}
The cavity consists of four mirrors in a bow-tie configuration symmetrical about a vertical line halfway between the mirrors (see Fig. \ref{fig:CavitySetupVRS}).
%
The two lower mirrors have 10 mm radius of curvature (ROC) and are separated by 11.1 mm. The two upper mirrors have 500 mm ROC and are separated by 20 mm.
The diagonal distance between a lower and upper mirror is 18.4 mm. The circulating light has an angle of incidence (AOI) of $16.4^{\circ}$ relative to the normal of every mirror.  
A slight non-planarity of the cavity gives rise to Pancharatnam phase~\cite{berr1987adia}-induced circular birefringence. The cavity modes are 90\% circularly polarized, with polarization modes split by 43 MHz, corresponding to a non-planarity of $6^{\circ}$.  

The cavity has a small waist located halfway between the lower mirrors and a larger waist halfway between the upper mirrors.
The waist sizes can be calculated with \textit{ABCD} matrices~\cite{sieg1986lase}.   Due to the non-zero AOI, the cavity is astigmatic, resulting in different waist sizes for the two transverse axes.  
The small waist has $1/e^2$ intensity radii of 12 $\mu$m horizontally and 11 $\mu$m  vertically.  The large waist has radii of 94 $\mu$m horizontally and 58 $\mu$m vertically.  The stability diagram of the resonator, derived from the \textit{ABCD} matrices, as a function of the separation of the lower mirrors (short arm), is shown in Fig. \ref{fig:CavityStability}(a).

The astigmatism of the cavity results in two distinct Gouy phases $\chi_{1,2}$ for horizontal and vertical transverse modes, respectively. The cavity length is tuned near the point $\chi_1 = 2 \chi_2$, where the mode frequencies exhibit a finite degeneracy $\nu_{m,n+2,q} = \nu_{m+1,n,q}$, where $m$ and $n$ are the horizontal  and vertical transverse mode indices, respectively, and $q$ is the longitudinal mode index. The transverse mode specturm is shown in Fig. \ref{fig:CavityStability}(b). We work slightly away from degeneracy, such that the TEM$_{02}$ and TEM$_{10}$ are split by 20 MHz. 


\subsection*{Detection}
Light is collected using a single photon counting module (SPCM). The total detection efficiency for a photon emitted from the outcoupling mirror of the cavity is measured to be about $\epsilon=0.21$, including the SPCM quantum efficiency $\eta_{det}=0.6$ and loss in optics $\eta_{optics}=0.35$. The maximum transmitted fraction through the outcoupling mirror at resonance on the TEM$_{00}$ mode for the empty cavity is $T_0=0.16$ due to losses and transmission through each of the four mirrors. The probe power impinging on the cavity is typically about 20 pW, with lower powers used for the higher principle quantum numbers to avoid shifts and broadening due to the build-up of ions or atoms in other Rydberg levels.

\subsection*{Atom Cloud and Transport}
We transport the atoms from the MOT to the resonator in an optical conveyor-belt (moving optical lattice) derived from a TA-amplified 784nm DBR laser, providing an axial trapping frequency of typically $90$ kHz. The conveyor belt has a waist $100 \mu$m in radius, located between the MOT and the lower resonator waist. 
The laser is narrowed by active electrical feedback from a PDH lock to a reference cavity with a 16 kHz linewidth.
To generate the moving lattice, we retro-reflect the laser through a double-passed tandem AOM, and rapidly tune the relative frequency of the two RF drives. 
The transport consists of acceleration at 980 m/s$^2$ to a cruising speed of 2 m/s, followed by a constant deceleration to rest in the cavity waist. 
The atomic cloud reaches the small (lower) cavity waist located 27 mm above the MOT after 16 ms of transport time.
The larger waist 36 mm above the MOT can also be reached.

The atomic cloud in the lattice has a temperature of 30 $\mu$K and a density distribution given approximately by a gaussian. The $1/e$ cloud radii are typicall 70 $\mu$m horizontally and 300 $\mu$m vertically. The peak density is typically about $3\times 10^{10}$ cm$^{-3}$. To vary the atomic density in the cavity, we adjust the total transport distance to sample different parts of the vertical density distribution. 

After loading the MOT, the atomic cloud is allowed to remain distributed over the $m_F$ magnetic sublevels of the F=2 ground state. The Rabi frequencies are different for atoms in different $m_F$ levels, reducing the effective couplings of the collective states.  A 1 G bias magnetic field is applied along the cavity axis.


\subsection*{Control laser Rabi Frequency}
\begin{figure}
\centering
  \includegraphics[width=70mm]{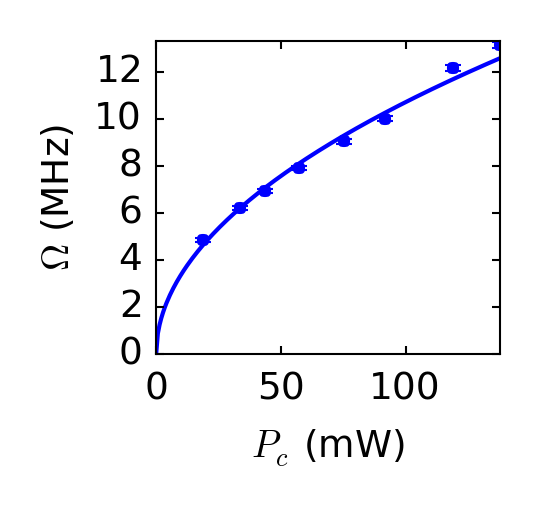}
  \caption{\textbf{Calibration of the control laser Rabi frequency}. We extract the blue (480 nm) control laser Rabi frequency $\Omega$ from fits to a complete three-level cavity EIT model. The Rabi frequency follows a square root dependence (fit: solid line) on the control laser power $P_c$.}
  \label{fig:BlueRabiCal}
\end{figure}
Figure \ref{fig:BlueRabiCal} shows the control laser Rabi frequency $\Omega$ obtained from fitting the cavity EIT transmission spectrum using the three-level model described later in this supplement. The data follow a square-root dependence on the laser power, as expected. The fit gives $\Omega\textrm{(MHz)}\approx$1.07 $\sqrt{P(\mathrm{mW})}$.

We use this calibration to provide the value of $\Omega$ from the laser power when studying the dependence of $\gamma_D$ and $d\delta_D/d\delta_c$ on $\theta$ in the main text, rather than determining $\Omega$ from each spectrum separately. 
The width of the dark polariton peak in the transmission spectrum is the main qualitative feature of the data that determines the value of $\Omega$ in the fit. Therefore, by using $\Omega$ from the control laser power we ensure that $\gamma_D$ and $\theta$ are independent in the comparison with the predicted $\gamma_D$. 

\subsection*{Measurement of Stray Electric Field}

\begin{figure}
  \includegraphics[width=80mm]{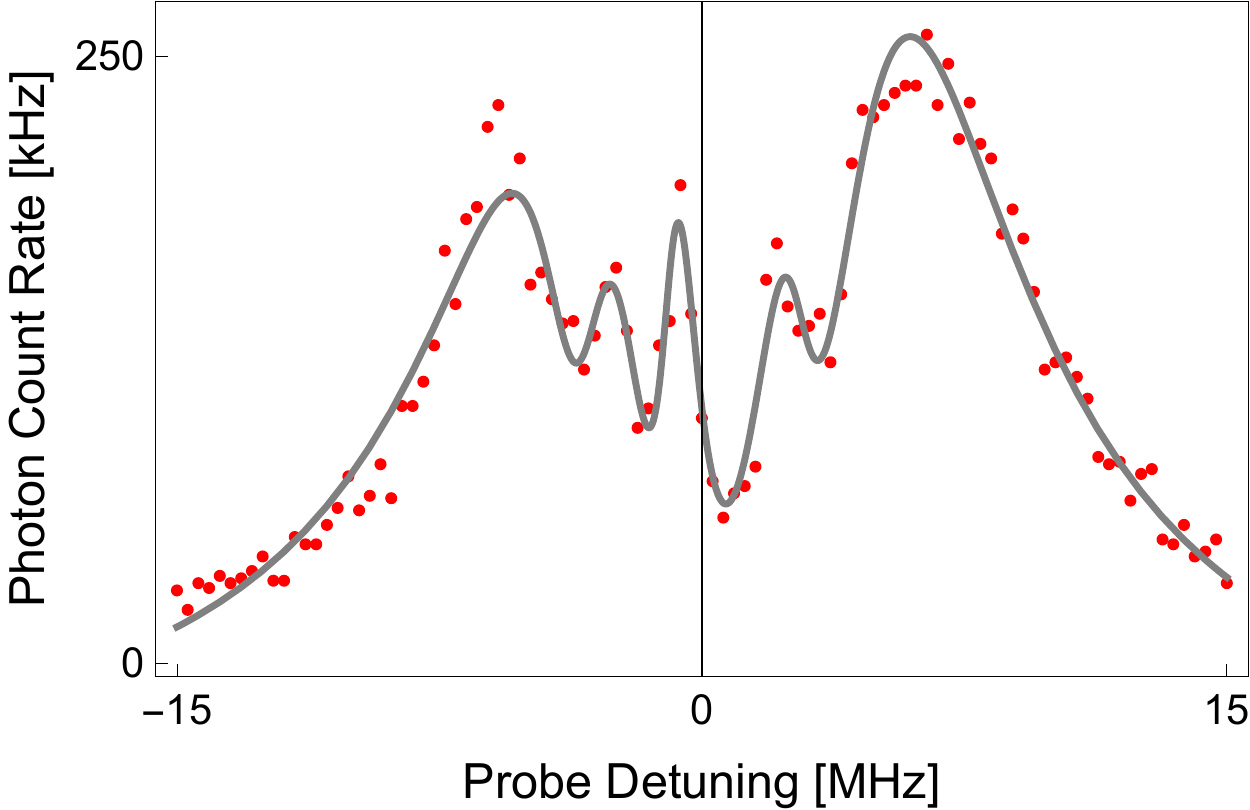}
  \caption{\textbf{E-field calibration through $38D_{5/2}$ cavity Rydberg EIT spectrum}. The cavity transmission is plotted versus probe laser frequency detuning from the ${}^{87}$Rb $5S_{1/2}(F=2)\rightarrow5P_{3/2}(F=3)$ transition, with the cavity tuned to the bare atomic resonance. The spectrum exhibits vacuum Rabi peaks, as well as $|m_j|=[1/2,3/2,5/2]$ EIT resonances.}
  \label{fig:38D}
\end{figure}

\begin{figure}[!tbp]
\includegraphics{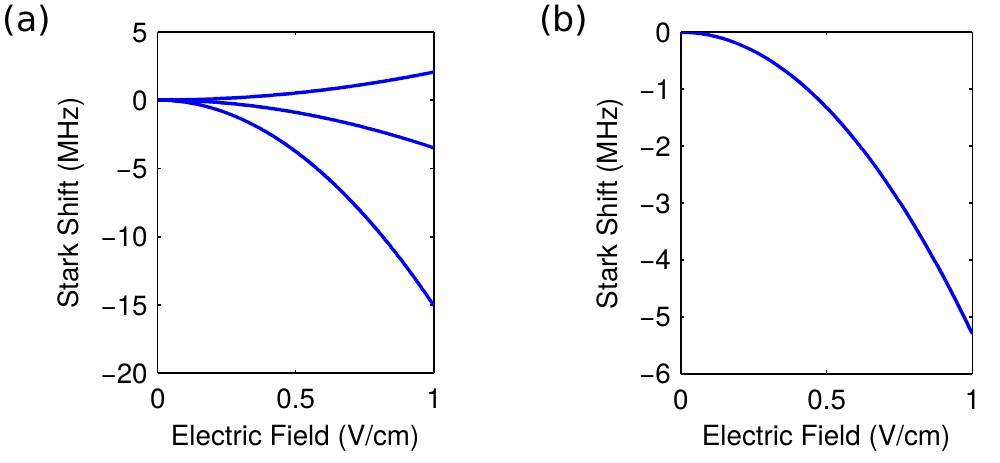}
  \caption{\textbf{Typical Calculated Stark Maps.} Energies of the 38D$_{5/2}$ (\textbf{Left}) and 40S (\textbf{Right}) $m_j$ sublevels versus applied electric field strength. For the 38D$_{5/2}$ state, the $|m_j|=1/2$ state has positive DC polarizability, reflected in its upward curvature, while the $|m_j|=3/2,5/2$ states have negative DC polarizability. The 40S states have negative polarizability. Note the difference in scale; the D-states generically exhibit larger DC polarizability than S-states for similar principal quantum number.\label{fig:stark}}
\end{figure}

We extract our local electric fields by measuring Rydberg EIT spectra in the 38D$_{5/2}$ state (Fig. \ref{fig:38D}), where a tensor DC stark splitting separates different $|m_j|$ levels. Combining the fitted locations of the different features with the computed tensor polarizability of 38D$_{5/2}$ enables us to extract the local electric field magnitude at the location of the atoms, though not its direction. The scalar and tensor polarizabilities, as well as full Stark maps (Fig. \ref{fig:stark}), of the relevant Rydberg levels are computed from Rydberg state wavefunctions obtained using Numerov integration. The measured field strength is 0.5 V/cm.

\subsection*{Buildup of Shelved Rydberg Atoms and Stray Ions}

\begin{figure}
\includegraphics{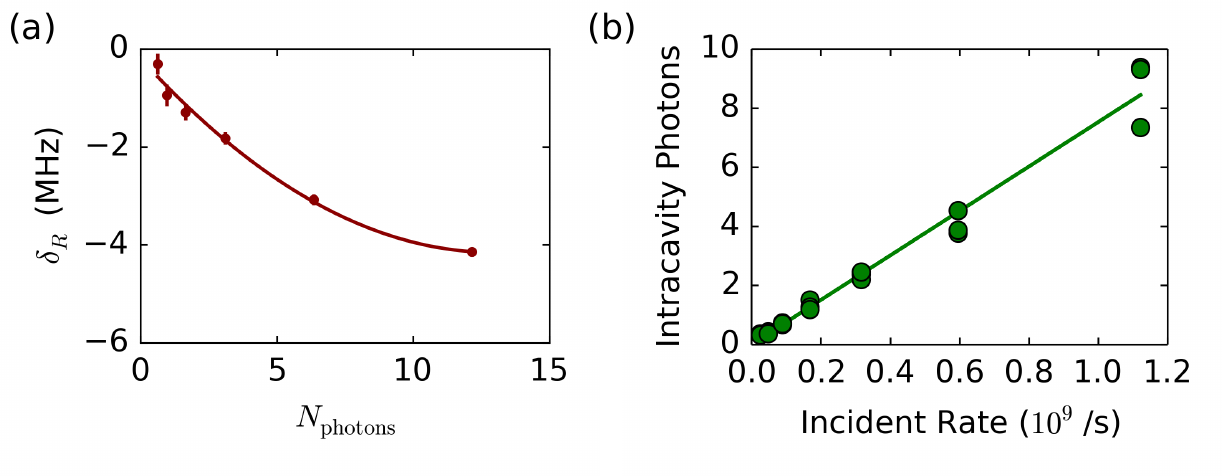}
  \caption{\textbf{Stray Ions and Shelved Rydberg Atoms.} 
 (a) We measure the shift of the Rydberg level vs probe power (quantified by the average intra-cavity photon number), for a 1 ms sweep $\pm$20MHz across the $60S$ Rydberg EIT feature. The atomic sample extends over more than 50 $\mu$m along the cavity axis, suggesting that polariton-polariton interactions are not the source of this frequency shift, but rather stray Rydbergs or ions. (b) 60S Rydberg EIT transmitted power vs. incident power, for extrapolated to zero-measurement time to avoid build-up of stray Rydbergs or ions. The observed linear relationship between the transmitted and incident power, for the probe on the low-power EIT resonance, indicates that there is indeed no observed polariton-polariton interaction.\label{fig:CavityFrequencyvsProbePower}
\label{fig:CavityPowervsProbePower}}
\end{figure}

The presence of ions or Rydberg atoms deposited in other principal quantum numbers through collisions~\cite{amth2007mode} can shift and broaden the original Rydberg level. 
To observe this effect, we increase the probe laser power beyond the range used in the main text, and plot the detuning $\delta_R$ of the Rydberg level at the dark polariton resonance observed during a 1 ms sweep of the probe laser, versus intra-cavity photon number (Fig. \ref{fig:CavityFrequencyvsProbePower}a).  The Rydberg level shifts downward with increasing intracavity photon number,  seeming to suggest a polariton-polariton nonlinearity. However, for the 60S Rydberg level used here, the van der Waals interactions are too weak to account for the observed shift in Fig. \ref{fig:CavityFrequencyvsProbePower}a. Furthermore, the sign of the shift indicates attractive interactions, while the nS-nS interaction is repulsive. On the other hand, the observed shift can be accounted for by the production of just a few ions.


Further evidence that the EIT frequency shift with intensity arises from ions or shelved Rydbergs rather than polariton-polariton scattering comes from a measurement (Fig. \ref{fig:CavityPowervsProbePower}b) of the peak transmitted power on the low intensity EIT resonance, versus probe intensity, measured only for \emph{short} probing time. Here photon counts for time $t< 50\,\mu$s at constant probe frequency are fit to a linear function and the $t=0$ value of the fit is used to compute the instantaneous intracavity photon number. The observed linearity rules out direct polariton-polariton interactions for these parameters, showing that buildup of background particles is the source of the shift in Fig. \ref{fig:CavityFrequencyvsProbePower}a.

\subsection{Decoherence Sources}
In the experiment, we observe an effective loss rate $\gamma_R$ from the collective Rydberg state at the dark polariton resonance of typically $\sim$300 kHz for the 40S Rydberg level. Laser frequency noise and $m_F$-dependent Zeeman shifts are the most likely dominant sources of loss, as all other leading sources contribute at the few kHz level. These other sources include: AC Stark shifts due to 1560 nm light used for cavity stabilization (including collective suppression of decoherence), Doppler broadening (also including collective suppression of decoherence), Penning ionization~\cite{amth2007mode}, and radiative decay~\cite{bete2009quas}. 

%
%

\section{Related Theory}
\subsection*{Cavity EIT Transmission Spectrum}
We derive the transmission spectrum in cavity EIT using perturbation theory for non-Hermitian Hamiltonians~\cite{cohe2004atom}. The non-Hermitian Hamiltonian description produces the same steady state as a master equation in the limit of weak driving~\cite{brec1999n, wu2008obse} (where the transmission spectrum is defined), and therefore we prefer it for the present purposes for simplicity.

The coupling of the cavity mode $\kt{C}$ to the collective atomic excitations $\kt{E}$ and $\kt{R}$ is described by the non-Hermitian Hamiltonian
\begin{eqnarray}
H_0/\hbar&=&(\deltacav-i\frac{\kappa}{2})\kt{C}\br{C}-i\frac{\Gamma}{2}\kt{E}\br{E} +(\deltaryd-i\frac{\gammaryd}{2})\kt{R}\br{R}+\label{eq:Hac}\\
	&&  +\frac{G}{2}\left(\kt{E}\br{C}+\kt{C}\br{E}\right) +\frac{\Omega}{2}(\kt{R}\br{E}+\kt{E}\br{R})\nonumber
\end{eqnarray}
To describe the probe laser, we include a fourth quantum state, $\kt{0}$, in which the cavity contains neither photons nor atomic excitations, and introduce the coupling to the probe laser, $V=q\widetilde{V}$, where $q$ is a small parameter and  
\begin{equation}
	\widetilde{V}= \kt{C}\br{0} + \kt{0}\br{C}
	\label{eqn:Vtilde}
\end{equation}
In the rotating frame with respect to the probe laser, the empty state has energy $\hbar\delta_L$ so that the total un-perturbed Hamiltonian is 
\begin{equation}
H_1 = \hbar\delta_L\kt{0}\br{0} + H_0
\end{equation}
The total perturbed Hamiltonian is 
\begin{equation}
H_t = H_1+ q\widetilde{V}
\end{equation}

For $q=0$, note that $\kt{0}$ is an eigenstate of $H_t$. Starting in $\kt{0}$ with $q=0$, we adiabatically increase $q$ and obtain the resulting eigenstate of $H_t$ using perturbation theory,
\begin{equation}
\kt{\psi}=\kt{0} + \frac{1}{\hbar\delta_L \mathds{1}-H_0}q\widetilde{V}\kt{0}+\mathcal{O}(q^2)
\end{equation}
where the inverse is taken in the $\left\{\kt{C},\kt{E},\kt{R}\right\}$ subspace. The fraction of light transmitted through the cavity is then given by 
\begin{equation}
T(\delta_L) = T_0\left(\frac{\kappa}{2}\right)^2\left|\br{C}\frac{1}{\delta_L \mathds{1}-H_0/\hbar}\kt{C}\right|^2,
\label{eqn:inverseTrans}
\end{equation}
where $T_0$ is the maximum transmission of light through the cavity in the absence of an atomic medium. For an ideal  symmetric two-mirror resonator, $T_0=1$. In our cavity, which has four non-identical mirrors, $T_0=0.16$ when collecting light from one of the two upper mirrors.

Evaluating the inverse of the resulting 3x3 matrix gives
\begin{equation} \label{eq:PertTheoryTransmissionSig}
T(\delta_L)=T_0\left(\frac{\kappa}{2}\right)^2\left|\frac{-4 \widetilde{\gamma}\widetilde{\Gamma} + \Omega^2}{G^2 \widetilde{\gamma}+\widetilde{\kappa}(-4 \widetilde{\gamma}\widetilde{\Gamma} + \Omega^2)}\right|^2,
\end{equation}

\noindent where $\widetilde{\kappa}\equiv-i \frac{\kappa}{2}+\delta_c-\delta_L$, $\widetilde{\Gamma}\equiv-i \frac{\Gamma}{2}-\delta_L$, $\widetilde{\gamma}\equiv-i \frac{\gamma_R}{2}+\delta_R-\delta_L$.

\subsection{EIT Window}
We study the EIT window in cavity EIT by looking at the height $T_D$ of the dark polariton resonance at $\delta_L=\deltadark$ as a function of the cavity detuning $\deltacav$. By considering an eigenvalue decomposition of the non-Hermitian Hamiltonian, we find
\begin{equation}
  T_{D}(\delta_c) \approx T_1 \left[\frac{1}{1+(\delta_c-\delta_R)^2/(\Gamma_w/2)^2}\right]^2
\end{equation}
where $T_1$ is the transmission when the cavity is at EIT resonance and 
\begin{equation}
\Gamma_w = \sqrt{\Gamma_{w1}^2 + \Gamma_{w2}^2}
\end{equation}
with
\begin{eqnarray}
\Gamma_{w1} &=& \frac{\Omega^2+G^2}{G}\sqrt{\frac{\kappa}{\Gamma}}\label{eq:Gammawone}\\
\Gamma_{w2} &=& \sqrt{ \frac{(\Omega^2+G^2)^2}{\Omega^2}\frac{\gamma_R}{\Gamma}-(\gamma_R-\kappa)^2}
\end{eqnarray}

We find $T_1$ from (\ref{eq:PertTheoryTransmissionSig}),
\begin{equation}
T_1 = T_0\left[1+\frac{\gamma_R}{\kappa}\frac{G^2}{\Omega^2+\gamma_R\Gamma}\right]^{-2}
\label{eq:Tone}
\end{equation}

It is instructive to relate the 
width in (\ref{eq:Gammawone}) to the free space EIT linewidth, given by $\Gamma_{\mathrm{EIT}}=\Omega^2/(\Gamma\sqrt{\mathrm{OD}})$ in the $\Omega<\Gamma$ limit, with $\mathrm{OD}$ the optical depth. To make the connection, view $\Tmax$ as a function of the resonant laser frequency $\delta_L=\deltadark\approx\deltacav\Omega^2/(\Omega^2+G^2)$. The corresponding width becomes $\Gamma_L=\Omega^2/(\Gamma\sqrt{\eta_c})$, where the collective cooperativity $\eta_c=G^2/(\kappa\Gamma)$ can also be identified as the cavity-enhanced optical density.

\subsection{Collective States in Cavity Rydberg EIT}
The collective states $\kt{E}$ and $\kt{R}$ are superpositions of different single-atom excitations. For a homogeneous control laser Rabi frequency $\Omega$ and arbitrary couplings $g_i$ of the $i^\mathrm{th}$ atom to the cavity mode, 

\begin{eqnarray}
G &=& \sqrt{\sum |g_i|^2} \nonumber \\
\ket{E} &=& \frac{1}{G} \sum g_i \ket{e_i} \nonumber \\
\ket{R} &=& \frac{1}{G}\sum g_i \ket{r_i},
\end{eqnarray}
where $\kt{e_i}$ ($\kt{r_i}$) is the state where the $i^\mathrm{th}$ atom is excited to the $\kt{e}$ ($\kt{r}$) level. 

These collective modes, and the associated three-state Hamiltonian (\ref{eq:Hac}), provide a useful approximate 3-state description of the system. However, effects that dephase the collective states create couplings to the larger bath of atomic excitations that are orthogonal to the ones given above. These orthogonal states rapidly decay by spontaneous emission through the $\kt{e}$ level. Such sources of dephasing include inhomogeneous broadening of the Rydberg level and Doppler broadening, both of which lead to non-uniform detunings from EIT resonance. While the inhomogeneity of the control laser intensity also creates couplings to this bath, we will see below that these couplings do not directly broaden the EIT dark state.

\subsubsection{Inhomogeneous Control Field}
Due to the finite waist size of the control laser, the control-laser Rabi frequency is not equal for all atoms. As a result, the Hamiltonian is no longer closed over the subspace spanned by $\{\kt{C}, \kt{E}, \kt{R}\}$. However, a dark state of the laser couplings still exists. The relevant collective Rydberg state is defined by the condition $H_{\Omega}\kt{R}=\Omega\kt{E}$, with the control laser coupling Hamiltonian given by
\begin{equation}
H_{\Omega} = \sum_i \omega_i \kt{r_i}\br{e_i} + \mathrm{h.c.}
\end{equation}
It follows that
\begin{eqnarray}
\Omega &=& \frac{G}{\sqrt{\sum_i |g_i/\omega_i|^2}}\label{eqn:OmegaInhom}\\
\kt{R} &=& \sum_i\frac{g_i/G}{\omega_i^*/\Omega} \kt{r_i}\label{eqn:Rinhom}
\end{eqnarray} 
The dark state is then given, as usual, by
\begin{equation}
\kt{D} = \frac{1}{\sqrt{\Omega^2+G^2}}\left(\Omega\kt{C} - G\kt{R}\right)
\label{eqn:Darkstate}
\end{equation}
Note that (\ref{eqn:Rinhom}) shows that the control laser must be wider in radius than the cavity mode at the location of the atoms. Otherwise, the atoms furthest from the laser beam have the largest contribution to $\kt{R}$. Numerical simulations of many atoms coupled to the cavity confirm that the inhomogeneous control laser does not produce significant loss in the dark state, and reproduce the coefficients in the wavefunction (\ref{eqn:Rinhom}). 

\subsection{Transmission Spectra for Two Cavity Modes}

For an ensemble of atoms coupled to a several optical modes, the phase and amplitude profile collectively written onto the atoms upon absorption of a resonator photon in a particular mode will  reflect the mode function of that mode. Collective states resulting from different cavity modes are therefore orthogonal in the limit of a uniform atomic density. For a two-mode cavity with nearly degenerate modes $1$ and $2$, this results in a 6-by-6 Hamiltonian in the basis \{$\ket{C_1}$, $\ket{E_1}$, $\ket{R_1}$, $\ket{C_2}$, $\ket{E_2}$, $\ket{R_2}$\}. Here $\ket{C_j}$ is a photon in optical mode $j$; $\ket{E_j}$ is a collective atomic excitation in the P-state, absorbed from mode $j$; and $\ket{R_j}$ is the corresponding collective atomic excitation in the Rydberg state. The resulting Hamiltonian is block diagonal:
\begin{equation}
H = 
\left( \begin{array}{cccccc}
-i\frac{\kappa_1}{2}  + \delta_{c,1} & G/2                                                                     & 0                                                                & 0                                                                        & 0                                                                        & 0 \\
G/2                                                                    & -i\frac{\Gamma}{2}  			 &  \Omega/2                                                & 0                                                                        & 0                                                                        & 0  \\
0                                                                       & \Omega/2                                                       & -i\frac{\gamma_R}{2}  +\delta_R & 0                                                                          &0                                                                        & 0 \\
0                                                                       &0                                                                       & 0                                                                 & -i\frac{\kappa_2}{2} + \delta_{c,2}   & G/2                                                                     &0 \\
0                                                                       & 0                                                                       & 0                                                   & G/2                                                                     & -i\frac{\Gamma}{2}      & \Omega/2 \\
0                                                                       & 0                                                                       & 0                                                                &0                                                                          & \Omega/2                   & -i\frac{\gamma_R}{2}  + \delta_R\\
\end{array} \right).
\end{equation}
Atomic-lensing-induced coupling between resonator modes due to a non-uniform atomic density can be modeled by non-zero couplings in the off-diagonal blocks, connecting $\kt{E_i}$ to $\kt{C_j}$ and $\kt{R_j}$ for $i\neq j$. 

To compute the transmission spectrum for a multi-mode cavity, we consider the coupling of the control laser to each cavity mode, so that the perturbative coupling (\ref{eqn:Vtilde}) becomes
\begin{equation}
\widetilde{V}=\sum_i a_i\kt{C_i}\br{0} + \mathrm{h.c.}
\end{equation}
with  $\sum_i|a_i|^2= 1$. For multi-mode experiments, we collect light in a multi-mode optical fiber. As the SPCM detects on multiple spatial modes, the detected signal is a sum of the transmitted \emph{powers}, and not the fields, washing out any interference between the modes. The resulting transmission spectrum is therefore
\begin{equation}
T(\delta_L) = \sum_i |a_i|^2 T^{(i)}(\delta_L)
\end{equation}
where $T^{(i)}$ is the transmission spectrum for the $i^{\mathrm{th}}$ mode given by Eqn. \eqref{eq:PertTheoryTransmissionSig} applied to that mode.

\section*{Monte Carlo Treatment of Transmission Spectra}

To validate our understanding of the dissipation channels arising from inhomogeneous broadening, we employ non-Hermitian perturbation theory for a large number of atoms coupled to a single mode resonator, and compute the transmission spectrum from the full Hamiltonian using (\ref{eqn:inverseTrans}).

The complete $N$ atom, $(2N+1)\times(2N+1)$ Hamiltonian is:

\[
\left(
\begin{array} {c |c c c | c c c}
-i \kappa/2&g_1/2&\cdots&g_n/2&0&\cdots&0\\ \hline
g_1/2&-i \Gamma/2&\cdots&0&\omega_1/2&\cdots&0\\
\vdots&\vdots&\ddots&\vdots&\vdots&\vdots&\vdots \\
g_n/2&0&\cdots&-i \Gamma/2&0&\cdots&\omega_n/2\\ \hline
0&\omega_1/2&\cdots&0&-i \Gamma_R/2+\delta_{R1}&\cdots&0\\
\vdots&\vdots&\vdots&\vdots&\vdots&\ddots&\vdots\\ 
0&0&\cdots&\omega_n/2&0&\cdots&-i \Gamma_R/2+\delta_{Rn}
\end{array}
\right)
\]

\noindent where the detunings $\delta_{Ri}$ include both an actual control laser detuning and the effects of any mechanisms that can create a detuning in the Rydberg state, and $\Gamma_R$ gives the homogeneous linewidth of the  Rydberg state. Atoms are assigned random locations within the cloud, and $g_i$ and $\omega_i$ are determined by the local amplitudes of the cavity mode and control field, respectively. The $g_i$ and $\omega_i$ are then rescaled to produce the desired total $G$ and $\Omega$, making those parameters independent of the number of atoms in the simulation. We use N=300 atoms, and run multiple trials for different randomized atom positions.

The effective loss rate $\gamma_R$ of the collective Rydberg state is obtained from the simulated spectra by fitting to the form (\ref{eq:PertTheoryTransmissionSig}) or from the maximum transmission (\ref{eq:Tone}).
This allows us to investigate the influence of Doppler broadening, electric field gradients, and inhomogeneous control laser Rabi frequency. We confirm the scaling $\gamma_R\propto\gamma_b^2\Gamma/\Omega^2$ through variation of parameters. Additionally, we confirm that inhomogeneous control laser intensity does not contribute significantly to the dark polariton loss.  However, it does present a non-zero contribution when, as is generally the case, $\kappa\neq\Gamma_R$. In this case, the dark state (\ref{eqn:Darkstate}) is not exact, due to a small P-state component, that couples to the bath of orthogonal collective modes when the control laser is non-uniform. For typical experimental parameters, this increases $\gamma_R$ by $\sim$ 2 kHz.

\begin{figure}
  \includegraphics[width=160mm]{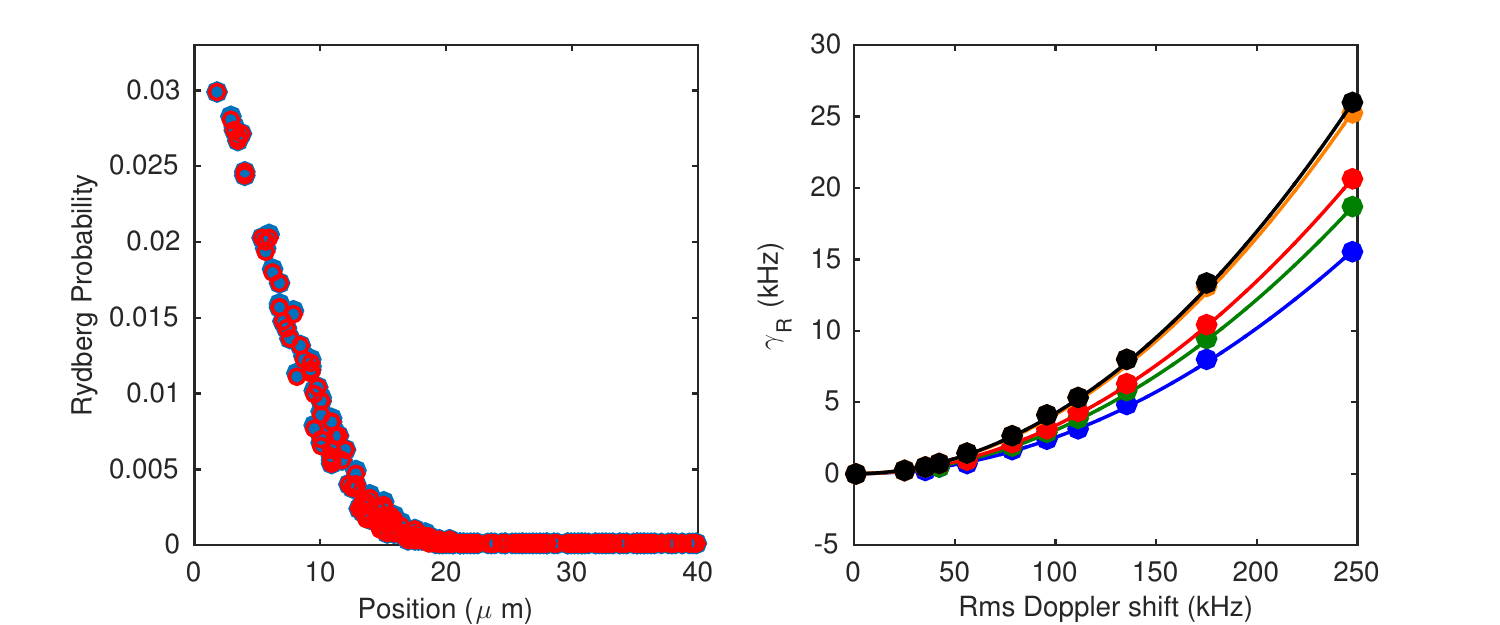}
  \caption{\textbf{Monte Carlo simulation of dark polariton wavefunction and Doppler broadening}. (\textbf{Left}) The probability of each atom to be in the Rydberg state is plotted versus radial position, referenced to the cavity mode axis. The blue points are generated from the dark state eigenvector of the simulation Hamiltonian, while the red, open circles are computed from (\ref{eqn:Darkstate}).  (\textbf{Right}) Simulated loss due to decoherence caused by atomic motion, for several realization of random atom locations.  The rms Doppler shift is quadratically suppressed. Curve: quadratic fit. In both plots we use $G$ (MHz) = 15, $\Omega$ (MHz) = 8, and other parameters set to the values used in the experiments. }

\end{figure}
 

\begin{thebibliography}{48}%
\makeatletter
\providecommand \@ifxundefined [1]{%
 \@ifx{#1\undefined}
}%
\providecommand \@ifnum [1]{%
 \ifnum #1\expandafter \@firstoftwo
 \else \expandafter \@secondoftwo
 \fi
}%
\providecommand \@ifx [1]{%
 \ifx #1\expandafter \@firstoftwo
 \else \expandafter \@secondoftwo
 \fi
}%
\providecommand \natexlab [1]{#1}%
\providecommand \enquote  [1]{``#1''}%
\providecommand \bibnamefont  [1]{#1}%
\providecommand \bibfnamefont [1]{#1}%
\providecommand \citenamefont [1]{#1}%
\providecommand \href@noop [0]{\@secondoftwo}%
\providecommand \href [0]{\begingroup \@sanitize@url \@href}%
\providecommand \@href[1]{\@@startlink{#1}\@@href}%
\providecommand \@@href[1]{\endgroup#1\@@endlink}%
\providecommand \@sanitize@url [0]{\catcode `\\12\catcode `\$12\catcode
  `\&12\catcode `\#12\catcode `\^12\catcode `\_12\catcode `\%12\relax}%
\providecommand \@@startlink[1]{}%
\providecommand \@@endlink[0]{}%
\providecommand \url  [0]{\begingroup\@sanitize@url \@url }%
\providecommand \@url [1]{\endgroup\@href {#1}{\urlprefix }}%
\providecommand \urlprefix  [0]{URL }%
\providecommand \Eprint [0]{\href }%
\providecommand \doibase [0]{http://dx.doi.org/}%
\providecommand \selectlanguage [0]{\@gobble}%
\providecommand \bibinfo  [0]{\@secondoftwo}%
\providecommand \bibfield  [0]{\@secondoftwo}%
\providecommand \translation [1]{[#1]}%
\providecommand \BibitemOpen [0]{}%
\providecommand \bibitemStop [0]{}%
\providecommand \bibitemNoStop [0]{.\EOS\space}%
\providecommand \EOS [0]{\spacefactor3000\relax}%
\providecommand \BibitemShut  [1]{\csname bibitem#1\endcsname}%
\let\auto@bib@innerbib\@empty
\bibitem [{\citenamefont {Balili}\ \emph {et~al.}(2007)\citenamefont {Balili},
  \citenamefont {Hartwell}, \citenamefont {Snoke}, \citenamefont {Pfeiffer},\
  and\ \citenamefont {West}}]{bali2007bose}%
  \BibitemOpen
  \bibfield  {author} {\bibinfo {author} {\bibfnamefont {R.}~\bibnamefont
  {Balili}}, \bibinfo {author} {\bibfnamefont {V.}~\bibnamefont {Hartwell}},
  \bibinfo {author} {\bibfnamefont {D.}~\bibnamefont {Snoke}}, \bibinfo
  {author} {\bibfnamefont {L.}~\bibnamefont {Pfeiffer}}, \ and\ \bibinfo
  {author} {\bibfnamefont {K.}~\bibnamefont {West}},\ }\href {\doibase
  10.1126/science.1140990} {\bibfield  {journal} {\bibinfo  {journal}
  {Science}\ }\textbf {\bibinfo {volume} {316}},\ \bibinfo {pages} {1007}
  (\bibinfo {year} {2007})}\BibitemShut {NoStop}%
\bibitem [{\citenamefont {Deng}\ \emph {et~al.}(2010)\citenamefont {Deng},
  \citenamefont {Haug},\ and\ \citenamefont {Yamamoto}}]{deng2010exci}%
  \BibitemOpen
  \bibfield  {author} {\bibinfo {author} {\bibfnamefont {H.}~\bibnamefont
  {Deng}}, \bibinfo {author} {\bibfnamefont {H.}~\bibnamefont {Haug}}, \ and\
  \bibinfo {author} {\bibfnamefont {Y.}~\bibnamefont {Yamamoto}},\ }\href@noop
  {} {\bibfield  {journal} {\bibinfo  {journal} {Rev. Mod. Phys.}\ }\textbf
  {\bibinfo {volume} {82}},\ \bibinfo {pages} {1489} (\bibinfo {year}
  {2010})}\BibitemShut {NoStop}%
\bibitem [{\citenamefont {Carusotto}\ and\ \citenamefont
  {Ciuti}(2013)}]{caru2013quan}%
  \BibitemOpen
  \bibfield  {author} {\bibinfo {author} {\bibfnamefont {I.}~\bibnamefont
  {Carusotto}}\ and\ \bibinfo {author} {\bibfnamefont {C.}~\bibnamefont
  {Ciuti}},\ }\href@noop {} {\bibfield  {journal} {\bibinfo  {journal} {Rev.
  Mod. Phys.}\ }\textbf {\bibinfo {volume} {85}},\ \bibinfo {pages} {299}
  (\bibinfo {year} {2013})}\BibitemShut {NoStop}%
\bibitem [{\citenamefont {Lukin}\ \emph {et~al.}(2001)\citenamefont {Lukin},
  \citenamefont {Fleischhauer}, \citenamefont {Cote}, \citenamefont {Duan},
  \citenamefont {Jaksch}, \citenamefont {Cirac},\ and\ \citenamefont
  {Zoller}}]{luki2001dipo}%
  \BibitemOpen
  \bibfield  {author} {\bibinfo {author} {\bibfnamefont {M.~D.}\ \bibnamefont
  {Lukin}}, \bibinfo {author} {\bibfnamefont {M.}~\bibnamefont {Fleischhauer}},
  \bibinfo {author} {\bibfnamefont {R.}~\bibnamefont {Cote}}, \bibinfo {author}
  {\bibfnamefont {L.~M.}\ \bibnamefont {Duan}}, \bibinfo {author}
  {\bibfnamefont {D.}~\bibnamefont {Jaksch}}, \bibinfo {author} {\bibfnamefont
  {J.~I.}\ \bibnamefont {Cirac}}, \ and\ \bibinfo {author} {\bibfnamefont
  {P.}~\bibnamefont {Zoller}},\ }\href {\doibase 10.1103/PhysRevLett.87.037901}
  {\bibfield  {journal} {\bibinfo  {journal} {Phys. Rev. Lett.}\ }\textbf
  {\bibinfo {volume} {87}},\ \bibinfo {pages} {037901} (\bibinfo {year}
  {2001})}\BibitemShut {NoStop}%
\bibitem [{\citenamefont {Pritchard}\ \emph {et~al.}(2010)\citenamefont
  {Pritchard}, \citenamefont {Maxwell}, \citenamefont {Gauguet}, \citenamefont
  {Weatherill}, \citenamefont {Jones},\ and\ \citenamefont
  {Adams}}]{prit2010coop}%
  \BibitemOpen
  \bibfield  {author} {\bibinfo {author} {\bibfnamefont {J.~D.}\ \bibnamefont
  {Pritchard}}, \bibinfo {author} {\bibfnamefont {D.}~\bibnamefont {Maxwell}},
  \bibinfo {author} {\bibfnamefont {A.}~\bibnamefont {Gauguet}}, \bibinfo
  {author} {\bibfnamefont {K.~J.}\ \bibnamefont {Weatherill}}, \bibinfo
  {author} {\bibfnamefont {M.~P.~A.}\ \bibnamefont {Jones}}, \ and\ \bibinfo
  {author} {\bibfnamefont {C.~S.}\ \bibnamefont {Adams}},\ }\href@noop {}
  {\bibfield  {journal} {\bibinfo  {journal} {Phys. Rev. Lett.}\ }\textbf
  {\bibinfo {volume} {105}},\ \bibinfo {pages} {193603} (\bibinfo {year}
  {2010})}\BibitemShut {NoStop}%
\bibitem [{\citenamefont {Gorshkov}\ \emph {et~al.}(2011)\citenamefont
  {Gorshkov}, \citenamefont {Otterbach}, \citenamefont {Fleischhauer},
  \citenamefont {Pohl},\ and\ \citenamefont {Lukin}}]{gors2011phot}%
  \BibitemOpen
  \bibfield  {author} {\bibinfo {author} {\bibfnamefont {A.~V.}\ \bibnamefont
  {Gorshkov}}, \bibinfo {author} {\bibfnamefont {J.}~\bibnamefont {Otterbach}},
  \bibinfo {author} {\bibfnamefont {M.}~\bibnamefont {Fleischhauer}}, \bibinfo
  {author} {\bibfnamefont {T.}~\bibnamefont {Pohl}}, \ and\ \bibinfo {author}
  {\bibfnamefont {M.~D.}\ \bibnamefont {Lukin}},\ }\href {\doibase
  10.1103/PhysRevLett.107.133602} {\bibfield  {journal} {\bibinfo  {journal}
  {Phys. Rev. Lett.}\ }\textbf {\bibinfo {volume} {107}},\ \bibinfo {pages}
  {133602} (\bibinfo {year} {2011})}\BibitemShut {NoStop}%
\bibitem [{\citenamefont {Dudin}\ and\ \citenamefont
  {Kuzmich}(2012)}]{dudi2012stro}%
  \BibitemOpen
  \bibfield  {author} {\bibinfo {author} {\bibfnamefont {Y.~O.}\ \bibnamefont
  {Dudin}}\ and\ \bibinfo {author} {\bibfnamefont {A.}~\bibnamefont
  {Kuzmich}},\ }\href@noop {} {\bibfield  {journal} {\bibinfo  {journal}
  {Science}\ }\textbf {\bibinfo {volume} {336}},\ \bibinfo {pages} {887}
  (\bibinfo {year} {2012})}\BibitemShut {NoStop}%
\bibitem [{\citenamefont {Peyronel}\ \emph {et~al.}(2012)\citenamefont
  {Peyronel}, \citenamefont {Firstenberg}, \citenamefont {Liang}, \citenamefont
  {Hofferberth}, \citenamefont {Gorshkov}, \citenamefont {Pohl}, \citenamefont
  {Lukin},\ and\ \citenamefont {Vuleti\'c}}]{peyr2012quan}%
  \BibitemOpen
  \bibfield  {author} {\bibinfo {author} {\bibfnamefont {T.}~\bibnamefont
  {Peyronel}}, \bibinfo {author} {\bibfnamefont {O.}~\bibnamefont
  {Firstenberg}}, \bibinfo {author} {\bibfnamefont {Q.~Y.}\ \bibnamefont
  {Liang}}, \bibinfo {author} {\bibfnamefont {S.}~\bibnamefont {Hofferberth}},
  \bibinfo {author} {\bibfnamefont {A.~V.}\ \bibnamefont {Gorshkov}}, \bibinfo
  {author} {\bibfnamefont {T.}~\bibnamefont {Pohl}}, \bibinfo {author}
  {\bibfnamefont {M.~D.}\ \bibnamefont {Lukin}}, \ and\ \bibinfo {author}
  {\bibfnamefont {V.}~\bibnamefont {Vuleti\'c}},\ }\href@noop {} {\bibfield
  {journal} {\bibinfo  {journal} {Nature}\ }\textbf {\bibinfo {volume} {488}},\
  \bibinfo {pages} {57} (\bibinfo {year} {2012})}\BibitemShut {NoStop}%
\bibitem [{\citenamefont {Maxwell}\ \emph {et~al.}(2013)\citenamefont
  {Maxwell}, \citenamefont {Szwer}, \citenamefont {Paredes-Barato},
  \citenamefont {Busche}, \citenamefont {Pritchard}, \citenamefont {Gauguet},
  \citenamefont {Weatherill}, \citenamefont {Jones},\ and\ \citenamefont
  {Adams}}]{maxw2013stor}%
  \BibitemOpen
  \bibfield  {author} {\bibinfo {author} {\bibfnamefont {D.}~\bibnamefont
  {Maxwell}}, \bibinfo {author} {\bibfnamefont {D.}~\bibnamefont {Szwer}},
  \bibinfo {author} {\bibfnamefont {D.}~\bibnamefont {Paredes-Barato}},
  \bibinfo {author} {\bibfnamefont {H.}~\bibnamefont {Busche}}, \bibinfo
  {author} {\bibfnamefont {J.}~\bibnamefont {Pritchard}}, \bibinfo {author}
  {\bibfnamefont {A.}~\bibnamefont {Gauguet}}, \bibinfo {author} {\bibfnamefont
  {K.}~\bibnamefont {Weatherill}}, \bibinfo {author} {\bibfnamefont
  {M.}~\bibnamefont {Jones}}, \ and\ \bibinfo {author} {\bibfnamefont
  {C.}~\bibnamefont {Adams}},\ }\href@noop {} {\bibfield  {journal} {\bibinfo
  {journal} {Phys. Rev. Lett.}\ }\textbf {\bibinfo {volume} {110}} (\bibinfo
  {year} {2013})}\BibitemShut {NoStop}%
\bibitem [{\citenamefont {Hofmann}\ \emph {et~al.}(2013)\citenamefont
  {Hofmann}, \citenamefont {G\"unter}, \citenamefont {Schempp}, \citenamefont
  {Robert-de Saint-Vincent}, \citenamefont {G\"arttner}, \citenamefont {Evers},
  \citenamefont {Whitlock},\ and\ \citenamefont
  {Weidem\"uller}}]{hofm2013sub-}%
  \BibitemOpen
  \bibfield  {author} {\bibinfo {author} {\bibfnamefont {C.~S.}\ \bibnamefont
  {Hofmann}}, \bibinfo {author} {\bibfnamefont {G.}~\bibnamefont {G\"unter}},
  \bibinfo {author} {\bibfnamefont {H.}~\bibnamefont {Schempp}}, \bibinfo
  {author} {\bibfnamefont {M.}~\bibnamefont {Robert-de Saint-Vincent}},
  \bibinfo {author} {\bibfnamefont {M.}~\bibnamefont {G\"arttner}}, \bibinfo
  {author} {\bibfnamefont {J.}~\bibnamefont {Evers}}, \bibinfo {author}
  {\bibfnamefont {S.}~\bibnamefont {Whitlock}}, \ and\ \bibinfo {author}
  {\bibfnamefont {M.}~\bibnamefont {Weidem\"uller}},\ }\href {\doibase
  10.1103/PhysRevLett.110.203601} {\bibfield  {journal} {\bibinfo  {journal}
  {Phys. Rev. Lett.}\ }\textbf {\bibinfo {volume} {110}},\ \bibinfo {pages}
  {203601} (\bibinfo {year} {2013})}\BibitemShut {NoStop}%
\bibitem [{\citenamefont {Firstenberg}\ \emph {et~al.}(2013)\citenamefont
  {Firstenberg}, \citenamefont {Peyronel}, \citenamefont {Liang}, \citenamefont
  {Gorshkov}, \citenamefont {Lukin},\ and\ \citenamefont
  {Vuleti\'c}}]{firs2013attr}%
  \BibitemOpen
  \bibfield  {author} {\bibinfo {author} {\bibfnamefont {O.}~\bibnamefont
  {Firstenberg}}, \bibinfo {author} {\bibfnamefont {T.}~\bibnamefont
  {Peyronel}}, \bibinfo {author} {\bibfnamefont {Q.~Y.}\ \bibnamefont {Liang}},
  \bibinfo {author} {\bibfnamefont {A.~V.}\ \bibnamefont {Gorshkov}}, \bibinfo
  {author} {\bibfnamefont {M.~D.}\ \bibnamefont {Lukin}}, \ and\ \bibinfo
  {author} {\bibfnamefont {V.}~\bibnamefont {Vuleti\'c}},\ }\href@noop {}
  {\bibfield  {journal} {\bibinfo  {journal} {Nature}\ }\textbf {\bibinfo
  {volume} {502}},\ \bibinfo {pages} {71} (\bibinfo {year} {2013})}\BibitemShut
  {NoStop}%
\bibitem [{\citenamefont {Bienias}\ \emph {et~al.}(2014)\citenamefont
  {Bienias}, \citenamefont {Choi}, \citenamefont {Firstenberg}, \citenamefont
  {Maghrebi}, \citenamefont {Gullans}, \citenamefont {Lukin}, \citenamefont
  {Gorshkov},\ and\ \citenamefont {B\"uchler}}]{bien2014scat}%
  \BibitemOpen
  \bibfield  {author} {\bibinfo {author} {\bibfnamefont {P.}~\bibnamefont
  {Bienias}}, \bibinfo {author} {\bibfnamefont {S.}~\bibnamefont {Choi}},
  \bibinfo {author} {\bibfnamefont {O.}~\bibnamefont {Firstenberg}}, \bibinfo
  {author} {\bibfnamefont {M.~F.}\ \bibnamefont {Maghrebi}}, \bibinfo {author}
  {\bibfnamefont {M.}~\bibnamefont {Gullans}}, \bibinfo {author} {\bibfnamefont
  {M.~D.}\ \bibnamefont {Lukin}}, \bibinfo {author} {\bibfnamefont {A.~V.}\
  \bibnamefont {Gorshkov}}, \ and\ \bibinfo {author} {\bibfnamefont {H.~P.}\
  \bibnamefont {B\"uchler}},\ }\href@noop {} {\bibfield  {journal} {\bibinfo
  {journal} {Phys. Rev. A}\ }\textbf {\bibinfo {volume} {90}},\ \bibinfo
  {pages} {053804} (\bibinfo {year} {2014})}\BibitemShut {NoStop}%
\bibitem [{\citenamefont {Grusdt}\ and\ \citenamefont
  {Fleischhauer}(2013)}]{grus2013frac}%
  \BibitemOpen
  \bibfield  {author} {\bibinfo {author} {\bibfnamefont {F.}~\bibnamefont
  {Grusdt}}\ and\ \bibinfo {author} {\bibfnamefont {M.}~\bibnamefont
  {Fleischhauer}},\ }\href@noop {} {\bibfield  {journal} {\bibinfo  {journal}
  {Phys. Rev. A}\ }\textbf {\bibinfo {volume} {87}},\ \bibinfo {pages} {043628}
  (\bibinfo {year} {2013})}\BibitemShut {NoStop}%
\bibitem [{\citenamefont {Hafezi}\ \emph {et~al.}(2013)\citenamefont {Hafezi},
  \citenamefont {Lukin},\ and\ \citenamefont {Taylor}}]{hafe2013non-}%
  \BibitemOpen
  \bibfield  {author} {\bibinfo {author} {\bibfnamefont {M.}~\bibnamefont
  {Hafezi}}, \bibinfo {author} {\bibfnamefont {M.~D.}\ \bibnamefont {Lukin}}, \
  and\ \bibinfo {author} {\bibfnamefont {J.~M.}\ \bibnamefont {Taylor}},\
  }\href@noop {} {\bibfield  {journal} {\bibinfo  {journal} {New J. Phys.}\
  }\textbf {\bibinfo {volume} {15}},\ \bibinfo {pages} {063001} (\bibinfo
  {year} {2013})}\BibitemShut {NoStop}%
\bibitem [{\citenamefont {Umucal{\i}lar}\ \emph {et~al.}(2014)\citenamefont
  {Umucal{\i}lar}, \citenamefont {Wouters},\ and\ \citenamefont
  {Carusotto}}]{umuc2014prob}%
  \BibitemOpen
  \bibfield  {author} {\bibinfo {author} {\bibfnamefont {R.~O.}\ \bibnamefont
  {Umucal{\i}lar}}, \bibinfo {author} {\bibfnamefont {M.}~\bibnamefont
  {Wouters}}, \ and\ \bibinfo {author} {\bibfnamefont {I.}~\bibnamefont
  {Carusotto}},\ }\href@noop {} {\bibfield  {journal} {\bibinfo  {journal}
  {Phys. Rev. A}\ }\textbf {\bibinfo {volume} {89}},\ \bibinfo {pages} {023803}
  (\bibinfo {year} {2014})}\BibitemShut {NoStop}%
\bibitem [{\citenamefont {Maghrebi}\ \emph {et~al.}(2015)\citenamefont
  {Maghrebi}, \citenamefont {Yao}, \citenamefont {Hafezi}, \citenamefont
  {Pohl}, \citenamefont {Firstenberg},\ and\ \citenamefont
  {Gorshkov}}]{magh2015frac}%
  \BibitemOpen
  \bibfield  {author} {\bibinfo {author} {\bibfnamefont {M.~F.}\ \bibnamefont
  {Maghrebi}}, \bibinfo {author} {\bibfnamefont {N.~Y.}\ \bibnamefont {Yao}},
  \bibinfo {author} {\bibfnamefont {M.}~\bibnamefont {Hafezi}}, \bibinfo
  {author} {\bibfnamefont {T.}~\bibnamefont {Pohl}}, \bibinfo {author}
  {\bibfnamefont {O.}~\bibnamefont {Firstenberg}}, \ and\ \bibinfo {author}
  {\bibfnamefont {A.~V.}\ \bibnamefont {Gorshkov}},\ }\href {\doibase
  10.1103/PhysRevA.91.033838} {\bibfield  {journal} {\bibinfo  {journal} {Phys.
  Rev. A}\ }\textbf {\bibinfo {volume} {91}},\ \bibinfo {pages} {033838}
  (\bibinfo {year} {2015})}\BibitemShut {NoStop}%
\bibitem [{\citenamefont {Sommer}\ \emph {et~al.}(2015)\citenamefont {Sommer},
  \citenamefont {B\"uchler},\ and\ \citenamefont {Simon}}]{somm2015quan}%
  \BibitemOpen
  \bibfield  {author} {\bibinfo {author} {\bibfnamefont {A.}~\bibnamefont
  {Sommer}}, \bibinfo {author} {\bibfnamefont {H.~P.}\ \bibnamefont
  {B\"uchler}}, \ and\ \bibinfo {author} {\bibfnamefont {J.}~\bibnamefont
  {Simon}},\ }\href@noop {} {\bibfield  {journal} {\bibinfo  {journal}
  {arXiv:1506.00341 [cond-mat.quant-gas]}\ } (\bibinfo {year}
  {2015})}\BibitemShut {NoStop}%
\bibitem [{\citenamefont {Chang}\ \emph {et~al.}(2008)\citenamefont {Chang},
  \citenamefont {Gritsev}, \citenamefont {Morigi}, \citenamefont {Vuleti\'c},
  \citenamefont {Lukin},\ and\ \citenamefont {Demler}}]{chan2008crys}%
  \BibitemOpen
  \bibfield  {author} {\bibinfo {author} {\bibfnamefont {D.~E.}\ \bibnamefont
  {Chang}}, \bibinfo {author} {\bibfnamefont {V.}~\bibnamefont {Gritsev}},
  \bibinfo {author} {\bibfnamefont {G.}~\bibnamefont {Morigi}}, \bibinfo
  {author} {\bibfnamefont {V.}~\bibnamefont {Vuleti\'c}}, \bibinfo {author}
  {\bibfnamefont {M.~D.}\ \bibnamefont {Lukin}}, \ and\ \bibinfo {author}
  {\bibfnamefont {E.~A.}\ \bibnamefont {Demler}},\ }\href {\doibase
  10.1038/nphys1074} {\bibfield  {journal} {\bibinfo  {journal} {Nature Phys.}\
  }\textbf {\bibinfo {volume} {4}},\ \bibinfo {pages} {884} (\bibinfo {year}
  {2008})}\BibitemShut {NoStop}%
\bibitem [{\citenamefont {Gopalakrishnan}\ \emph {et~al.}(2009)\citenamefont
  {Gopalakrishnan}, \citenamefont {Lev},\ and\ \citenamefont
  {Goldbart}}]{gopa2009emer}%
  \BibitemOpen
  \bibfield  {author} {\bibinfo {author} {\bibfnamefont {S.}~\bibnamefont
  {Gopalakrishnan}}, \bibinfo {author} {\bibfnamefont {B.~L.}\ \bibnamefont
  {Lev}}, \ and\ \bibinfo {author} {\bibfnamefont {P.~M.}\ \bibnamefont
  {Goldbart}},\ }\href@noop {} {\bibfield  {journal} {\bibinfo  {journal}
  {Nature Phys.}\ }\textbf {\bibinfo {volume} {5}},\ \bibinfo {pages} {845}
  (\bibinfo {year} {2009})}\BibitemShut {NoStop}%
\bibitem [{\citenamefont {Otterbach}\ \emph {et~al.}(2013)\citenamefont
  {Otterbach}, \citenamefont {Moos}, \citenamefont {Muth},\ and\ \citenamefont
  {Fleischhauer}}]{otte2013wign}%
  \BibitemOpen
  \bibfield  {author} {\bibinfo {author} {\bibfnamefont {J.}~\bibnamefont
  {Otterbach}}, \bibinfo {author} {\bibfnamefont {M.}~\bibnamefont {Moos}},
  \bibinfo {author} {\bibfnamefont {D.}~\bibnamefont {Muth}}, \ and\ \bibinfo
  {author} {\bibfnamefont {M.}~\bibnamefont {Fleischhauer}},\ }\href@noop {}
  {\bibfield  {journal} {\bibinfo  {journal} {Phys. Rev. Lett.}\ }\textbf
  {\bibinfo {volume} {111}},\ \bibinfo {pages} {113001} (\bibinfo {year}
  {2013})}\BibitemShut {NoStop}%
\bibitem [{\citenamefont {Koll\'ar}\ \emph {et~al.}(2015)\citenamefont
  {Koll\'ar}, \citenamefont {Papageorge}, \citenamefont {Baumann},
  \citenamefont {Armen},\ and\ \citenamefont {Lev}}]{koll2015adju}%
  \BibitemOpen
  \bibfield  {author} {\bibinfo {author} {\bibfnamefont {A.~J.}\ \bibnamefont
  {Koll\'ar}}, \bibinfo {author} {\bibfnamefont {A.~T.}\ \bibnamefont
  {Papageorge}}, \bibinfo {author} {\bibfnamefont {K.}~\bibnamefont {Baumann}},
  \bibinfo {author} {\bibfnamefont {M.~A.}\ \bibnamefont {Armen}}, \ and\
  \bibinfo {author} {\bibfnamefont {B.~L.}\ \bibnamefont {Lev}},\ }\href@noop
  {} {\bibfield  {journal} {\bibinfo  {journal} {New. J Phys.}\ }\textbf
  {\bibinfo {volume} {17}},\ \bibinfo {pages} {043012} (\bibinfo {year}
  {2015})}\BibitemShut {NoStop}%
\bibitem [{\citenamefont {Klaers}\ \emph {et~al.}(2010)\citenamefont {Klaers},
  \citenamefont {Schmitt}, \citenamefont {Vewinger},\ and\ \citenamefont
  {Weitz}}]{klae2010bose}%
  \BibitemOpen
  \bibfield  {author} {\bibinfo {author} {\bibfnamefont {J.}~\bibnamefont
  {Klaers}}, \bibinfo {author} {\bibfnamefont {J.}~\bibnamefont {Schmitt}},
  \bibinfo {author} {\bibfnamefont {F.}~\bibnamefont {Vewinger}}, \ and\
  \bibinfo {author} {\bibfnamefont {M.}~\bibnamefont {Weitz}},\ }\href@noop {}
  {\bibfield  {journal} {\bibinfo  {journal} {Nature}\ }\textbf {\bibinfo
  {volume} {468}},\ \bibinfo {pages} {545} (\bibinfo {year}
  {2010})}\BibitemShut {NoStop}%
\bibitem [{\citenamefont {Sommer}\ and\ \citenamefont
  {Simon}(2015)}]{somm2015engi}%
  \BibitemOpen
  \bibfield  {author} {\bibinfo {author} {\bibfnamefont {A.}~\bibnamefont
  {Sommer}}\ and\ \bibinfo {author} {\bibfnamefont {J.}~\bibnamefont {Simon}},\
  }\href@noop {} {\bibfield  {journal} {\bibinfo  {journal} {arXiv:1511.00595
  [cond-mat, physics:physics, physics:quant-ph]}\ } (\bibinfo {year}
  {2015})}\BibitemShut {NoStop}%
\bibitem [{\citenamefont {Parigi}\ \emph {et~al.}(2012)\citenamefont {Parigi},
  \citenamefont {Bimbard}, \citenamefont {Stanojevic}, \citenamefont
  {Hilliard}, \citenamefont {Nogrette}, \citenamefont {Tualle-Brouri},
  \citenamefont {Ourjoumtsev},\ and\ \citenamefont {Grangier}}]{pari2012obse}%
  \BibitemOpen
  \bibfield  {author} {\bibinfo {author} {\bibfnamefont {V.}~\bibnamefont
  {Parigi}}, \bibinfo {author} {\bibfnamefont {E.}~\bibnamefont {Bimbard}},
  \bibinfo {author} {\bibfnamefont {J.}~\bibnamefont {Stanojevic}}, \bibinfo
  {author} {\bibfnamefont {A.~J.}\ \bibnamefont {Hilliard}}, \bibinfo {author}
  {\bibfnamefont {F.}~\bibnamefont {Nogrette}}, \bibinfo {author}
  {\bibfnamefont {R.}~\bibnamefont {Tualle-Brouri}}, \bibinfo {author}
  {\bibfnamefont {A.}~\bibnamefont {Ourjoumtsev}}, \ and\ \bibinfo {author}
  {\bibfnamefont {P.}~\bibnamefont {Grangier}},\ }\href {\doibase
  10.1103/PhysRevLett.109.233602} {\bibfield  {journal} {\bibinfo  {journal}
  {Phys. Rev. Lett.}\ }\textbf {\bibinfo {volume} {109}},\ \bibinfo {pages}
  {233602} (\bibinfo {year} {2012})}\BibitemShut {NoStop}%
\bibitem [{\citenamefont {Stanojevic}\ \emph {et~al.}(2013)\citenamefont
  {Stanojevic}, \citenamefont {Parigi}, \citenamefont {Bimbard}, \citenamefont
  {Ourjoumtsev},\ and\ \citenamefont {Grangier}}]{stan2013disp}%
  \BibitemOpen
  \bibfield  {author} {\bibinfo {author} {\bibfnamefont {J.}~\bibnamefont
  {Stanojevic}}, \bibinfo {author} {\bibfnamefont {V.}~\bibnamefont {Parigi}},
  \bibinfo {author} {\bibfnamefont {E.}~\bibnamefont {Bimbard}}, \bibinfo
  {author} {\bibfnamefont {A.}~\bibnamefont {Ourjoumtsev}}, \ and\ \bibinfo
  {author} {\bibfnamefont {P.}~\bibnamefont {Grangier}},\ }\href {\doibase
  10.1103/PhysRevA.88.053845} {\bibfield  {journal} {\bibinfo  {journal} {Phys.
  Rev. A}\ }\textbf {\bibinfo {volume} {88}},\ \bibinfo {pages} {053845}
  (\bibinfo {year} {2013})}\BibitemShut {NoStop}%
\bibitem [{\citenamefont {Lukin}(2003)}]{luki2003coll}%
  \BibitemOpen
  \bibfield  {author} {\bibinfo {author} {\bibfnamefont {M.~D.}\ \bibnamefont
  {Lukin}},\ }\href {\doibase 10.1103/RevModPhys.75.457} {\bibfield  {journal}
  {\bibinfo  {journal} {Rev. Mod. Phys.}\ }\textbf {\bibinfo {volume} {75}},\
  \bibinfo {pages} {457} (\bibinfo {year} {2003})}\BibitemShut {NoStop}%
\bibitem [{\citenamefont {Fleischhauer}\ and\ \citenamefont
  {Lukin}(2000)}]{flei2000dark}%
  \BibitemOpen
  \bibfield  {author} {\bibinfo {author} {\bibfnamefont {M.}~\bibnamefont
  {Fleischhauer}}\ and\ \bibinfo {author} {\bibfnamefont {M.~D.}\ \bibnamefont
  {Lukin}},\ }\href@noop {} {\bibfield  {journal} {\bibinfo  {journal} {Phys.
  Rev. Lett.}\ }\textbf {\bibinfo {volume} {84}},\ \bibinfo {pages} {5094}
  (\bibinfo {year} {2000})}\BibitemShut {NoStop}%
\bibitem [{\citenamefont {Lukin}\ \emph {et~al.}(1998)\citenamefont {Lukin},
  \citenamefont {Fleischhauer}, \citenamefont {Scully},\ and\ \citenamefont
  {Velichansky}}]{luki1998intr}%
  \BibitemOpen
  \bibfield  {author} {\bibinfo {author} {\bibfnamefont {M.~D.}\ \bibnamefont
  {Lukin}}, \bibinfo {author} {\bibfnamefont {M.}~\bibnamefont {Fleischhauer}},
  \bibinfo {author} {\bibfnamefont {M.~O.}\ \bibnamefont {Scully}}, \ and\
  \bibinfo {author} {\bibfnamefont {V.~L.}\ \bibnamefont {Velichansky}},\
  }\href {\doibase 10.1364/OL.23.000295} {\bibfield  {journal} {\bibinfo
  {journal} {Opt. Lett}\ }\textbf {\bibinfo {volume} {23}},\ \bibinfo {pages}
  {295} (\bibinfo {year} {1998})}\BibitemShut {NoStop}%
\bibitem [{\citenamefont {Wang}\ \emph {et~al.}(2000)\citenamefont {Wang},
  \citenamefont {Goorskey}, \citenamefont {Burkett},\ and\ \citenamefont
  {Xiao}}]{wang2000cavi}%
  \BibitemOpen
  \bibfield  {author} {\bibinfo {author} {\bibfnamefont {H.}~\bibnamefont
  {Wang}}, \bibinfo {author} {\bibfnamefont {D.~J.}\ \bibnamefont {Goorskey}},
  \bibinfo {author} {\bibfnamefont {W.~H.}\ \bibnamefont {Burkett}}, \ and\
  \bibinfo {author} {\bibfnamefont {M.}~\bibnamefont {Xiao}},\ }\href {\doibase
  10.1364/OL.25.001732} {\bibfield  {journal} {\bibinfo  {journal} {Opt. Lett}\
  }\textbf {\bibinfo {volume} {25}},\ \bibinfo {pages} {1732} (\bibinfo {year}
  {2000})}\BibitemShut {NoStop}%
\bibitem [{\citenamefont {Hernandez}\ \emph {et~al.}(2007)\citenamefont
  {Hernandez}, \citenamefont {Zhang},\ and\ \citenamefont
  {Zhu}}]{hern2007vacu}%
  \BibitemOpen
  \bibfield  {author} {\bibinfo {author} {\bibfnamefont {G.}~\bibnamefont
  {Hernandez}}, \bibinfo {author} {\bibfnamefont {J.}~\bibnamefont {Zhang}}, \
  and\ \bibinfo {author} {\bibfnamefont {Y.}~\bibnamefont {Zhu}},\ }\href
  {\doibase 10.1103/PhysRevA.76.053814} {\bibfield  {journal} {\bibinfo
  {journal} {Phys. Rev. A}\ }\textbf {\bibinfo {volume} {76}},\ \bibinfo
  {pages} {053814} (\bibinfo {year} {2007})}\BibitemShut {NoStop}%
\bibitem [{\citenamefont {Wu}\ \emph {et~al.}(2008)\citenamefont {Wu},
  \citenamefont {Gea-Banacloche},\ and\ \citenamefont {Xiao}}]{wu2008obse}%
  \BibitemOpen
  \bibfield  {author} {\bibinfo {author} {\bibfnamefont {H.}~\bibnamefont
  {Wu}}, \bibinfo {author} {\bibfnamefont {J.}~\bibnamefont {Gea-Banacloche}},
  \ and\ \bibinfo {author} {\bibfnamefont {M.}~\bibnamefont {Xiao}},\ }\href
  {\doibase 10.1103/PhysRevLett.100.173602} {\bibfield  {journal} {\bibinfo
  {journal} {Phys. Rev. Lett.}\ }\textbf {\bibinfo {volume} {100}},\ \bibinfo
  {pages} {173602} (\bibinfo {year} {2008})}\BibitemShut {NoStop}%
\bibitem [{\citenamefont {M\"ucke}\ \emph {et~al.}(2010)\citenamefont
  {M\"ucke}, \citenamefont {Figueroa}, \citenamefont {Bochmann}, \citenamefont
  {Hahn}, \citenamefont {Murr}, \citenamefont {Ritter}, \citenamefont
  {Villas-Boas},\ and\ \citenamefont {Rempe}}]{mu2010elec}%
  \BibitemOpen
  \bibfield  {author} {\bibinfo {author} {\bibfnamefont {M.}~\bibnamefont
  {M\"ucke}}, \bibinfo {author} {\bibfnamefont {E.}~\bibnamefont {Figueroa}},
  \bibinfo {author} {\bibfnamefont {J.}~\bibnamefont {Bochmann}}, \bibinfo
  {author} {\bibfnamefont {C.}~\bibnamefont {Hahn}}, \bibinfo {author}
  {\bibfnamefont {K.}~\bibnamefont {Murr}}, \bibinfo {author} {\bibfnamefont
  {S.}~\bibnamefont {Ritter}}, \bibinfo {author} {\bibfnamefont {C.~J.}\
  \bibnamefont {Villas-Boas}}, \ and\ \bibinfo {author} {\bibfnamefont
  {G.}~\bibnamefont {Rempe}},\ }\href {\doibase doi:10.1038/nature09093}
  {\bibfield  {journal} {\bibinfo  {journal} {Nature}\ }\textbf {\bibinfo
  {volume} {465}},\ \bibinfo {pages} {755} (\bibinfo {year}
  {2010})}\BibitemShut {NoStop}%
\bibitem [{\citenamefont {Kampschulte}\ \emph {et~al.}(2014)\citenamefont
  {Kampschulte}, \citenamefont {Alt}, \citenamefont {Manz}, \citenamefont
  {Martinez-Dorantes}, \citenamefont {Reimann}, \citenamefont {Yoon},
  \citenamefont {Meschede}, \citenamefont {Bienert},\ and\ \citenamefont
  {Morigi}}]{kamp2014elec}%
  \BibitemOpen
  \bibfield  {author} {\bibinfo {author} {\bibfnamefont {T.}~\bibnamefont
  {Kampschulte}}, \bibinfo {author} {\bibfnamefont {W.}~\bibnamefont {Alt}},
  \bibinfo {author} {\bibfnamefont {S.}~\bibnamefont {Manz}}, \bibinfo {author}
  {\bibfnamefont {M.}~\bibnamefont {Martinez-Dorantes}}, \bibinfo {author}
  {\bibfnamefont {R.}~\bibnamefont {Reimann}}, \bibinfo {author} {\bibfnamefont
  {S.}~\bibnamefont {Yoon}}, \bibinfo {author} {\bibfnamefont {D.}~\bibnamefont
  {Meschede}}, \bibinfo {author} {\bibfnamefont {M.}~\bibnamefont {Bienert}}, \
  and\ \bibinfo {author} {\bibfnamefont {G.}~\bibnamefont {Morigi}},\ }\href
  {\doibase 10.1103/PhysRevA.89.033404} {\bibfield  {journal} {\bibinfo
  {journal} {Phys. Rev. A}\ }\textbf {\bibinfo {volume} {89}},\ \bibinfo
  {pages} {033404} (\bibinfo {year} {2014})}\BibitemShut {NoStop}%
\bibitem [{si()}]{si}%
  \BibitemOpen
  \href@noop {} {\ }\bibinfo {note} {See Supplemental Material at http:// for
  additional details on the experimental implementation, theoretical
  description, and numerical simulations.}\BibitemShut {Stop}%
\bibitem [{\citenamefont {Tauschinsky}\ \emph {et~al.}(2010)\citenamefont
  {Tauschinsky}, \citenamefont {Thijssen}, \citenamefont {Whitlock},
  \citenamefont {van Linden van~den Heuvell},\ and\ \citenamefont
  {Spreeuw}}]{taus2010spat}%
  \BibitemOpen
  \bibfield  {author} {\bibinfo {author} {\bibfnamefont {A.}~\bibnamefont
  {Tauschinsky}}, \bibinfo {author} {\bibfnamefont {R.~M.~T.}\ \bibnamefont
  {Thijssen}}, \bibinfo {author} {\bibfnamefont {S.}~\bibnamefont {Whitlock}},
  \bibinfo {author} {\bibfnamefont {H.~B.}\ \bibnamefont {van Linden van~den
  Heuvell}}, \ and\ \bibinfo {author} {\bibfnamefont {R.~J.~C.}\ \bibnamefont
  {Spreeuw}},\ }\href {\doibase 10.1103/PhysRevA.81.063411} {\bibfield
  {journal} {\bibinfo  {journal} {Phys. Rev. A}\ }\textbf {\bibinfo {volume}
  {81}},\ \bibinfo {pages} {063411} (\bibinfo {year} {2010})}\BibitemShut
  {NoStop}%
\bibitem [{\citenamefont {Hattermann}\ \emph {et~al.}(2012)\citenamefont
  {Hattermann}, \citenamefont {Mack}, \citenamefont {Karlewski}, \citenamefont
  {Jessen}, \citenamefont {Cano},\ and\ \citenamefont
  {Fort\'agh}}]{hatt2012detr}%
  \BibitemOpen
  \bibfield  {author} {\bibinfo {author} {\bibfnamefont {H.}~\bibnamefont
  {Hattermann}}, \bibinfo {author} {\bibfnamefont {M.}~\bibnamefont {Mack}},
  \bibinfo {author} {\bibfnamefont {F.}~\bibnamefont {Karlewski}}, \bibinfo
  {author} {\bibfnamefont {F.}~\bibnamefont {Jessen}}, \bibinfo {author}
  {\bibfnamefont {D.}~\bibnamefont {Cano}}, \ and\ \bibinfo {author}
  {\bibfnamefont {J.}~\bibnamefont {Fort\'agh}},\ }\href {\doibase
  10.1103/PhysRevA.86.022511} {\bibfield  {journal} {\bibinfo  {journal} {Phys.
  Rev. A}\ }\textbf {\bibinfo {volume} {86}},\ \bibinfo {pages} {022511}
  (\bibinfo {year} {2012})}\BibitemShut {NoStop}%
\bibitem [{\citenamefont {Hogan}\ \emph {et~al.}(2012)\citenamefont {Hogan},
  \citenamefont {Agner}, \citenamefont {Merkt}, \citenamefont {Thiele},
  \citenamefont {Filipp},\ and\ \citenamefont {Wallraff}}]{hoga2012driv}%
  \BibitemOpen
  \bibfield  {author} {\bibinfo {author} {\bibfnamefont {S.~D.}\ \bibnamefont
  {Hogan}}, \bibinfo {author} {\bibfnamefont {J.~A.}\ \bibnamefont {Agner}},
  \bibinfo {author} {\bibfnamefont {F.}~\bibnamefont {Merkt}}, \bibinfo
  {author} {\bibfnamefont {T.}~\bibnamefont {Thiele}}, \bibinfo {author}
  {\bibfnamefont {S.}~\bibnamefont {Filipp}}, \ and\ \bibinfo {author}
  {\bibfnamefont {A.}~\bibnamefont {Wallraff}},\ }\href {\doibase
  10.1103/PhysRevLett.108.063004} {\bibfield  {journal} {\bibinfo  {journal}
  {Phys. Rev. Lett.}\ }\textbf {\bibinfo {volume} {108}},\ \bibinfo {pages}
  {063004} (\bibinfo {year} {2012})}\BibitemShut {NoStop}%
\bibitem [{\citenamefont {Fleischhauer}\ \emph {et~al.}(2005)\citenamefont
  {Fleischhauer}, \citenamefont {Imamoglu},\ and\ \citenamefont
  {Marangos}}]{flei2005elec}%
  \BibitemOpen
  \bibfield  {author} {\bibinfo {author} {\bibfnamefont {M.}~\bibnamefont
  {Fleischhauer}}, \bibinfo {author} {\bibfnamefont {A.}~\bibnamefont
  {Imamoglu}}, \ and\ \bibinfo {author} {\bibfnamefont {J.~P.}\ \bibnamefont
  {Marangos}},\ }\href {\doibase 10.1103/RevModPhys.77.633} {\bibfield
  {journal} {\bibinfo  {journal} {Rev. Mod. Phys.}\ }\textbf {\bibinfo {volume}
  {77}},\ \bibinfo {pages} {633} (\bibinfo {year} {2005})}\BibitemShut
  {NoStop}%
\bibitem [{\citenamefont {Honer}\ \emph {et~al.}(2011)\citenamefont {Honer},
  \citenamefont {L\"ow}, \citenamefont {Weimer}, \citenamefont {Pfau},\ and\
  \citenamefont {B\"uchler}}]{hone2011arti}%
  \BibitemOpen
  \bibfield  {author} {\bibinfo {author} {\bibfnamefont {J.}~\bibnamefont
  {Honer}}, \bibinfo {author} {\bibfnamefont {R.}~\bibnamefont {L\"ow}},
  \bibinfo {author} {\bibfnamefont {H.}~\bibnamefont {Weimer}}, \bibinfo
  {author} {\bibfnamefont {T.}~\bibnamefont {Pfau}}, \ and\ \bibinfo {author}
  {\bibfnamefont {H.~P.}\ \bibnamefont {B\"uchler}},\ }\href {\doibase
  10.1103/PhysRevLett.107.093601} {\bibfield  {journal} {\bibinfo  {journal}
  {Phys. Rev. Lett.}\ }\textbf {\bibinfo {volume} {107}},\ \bibinfo {pages}
  {093601} (\bibinfo {year} {2011})}\BibitemShut {NoStop}%
\bibitem [{\citenamefont {Gorniaczyk}\ \emph {et~al.}(2014)\citenamefont
  {Gorniaczyk}, \citenamefont {Tresp}, \citenamefont {Schmidt}, \citenamefont
  {Fedder},\ and\ \citenamefont {Hofferberth}}]{gorn2014sing}%
  \BibitemOpen
  \bibfield  {author} {\bibinfo {author} {\bibfnamefont {H.}~\bibnamefont
  {Gorniaczyk}}, \bibinfo {author} {\bibfnamefont {C.}~\bibnamefont {Tresp}},
  \bibinfo {author} {\bibfnamefont {J.}~\bibnamefont {Schmidt}}, \bibinfo
  {author} {\bibfnamefont {H.}~\bibnamefont {Fedder}}, \ and\ \bibinfo {author}
  {\bibfnamefont {S.}~\bibnamefont {Hofferberth}},\ }\href {\doibase
  10.1103/PhysRevLett.113.053601} {\bibfield  {journal} {\bibinfo  {journal}
  {Phys. Rev. Lett.}\ }\textbf {\bibinfo {volume} {113}},\ \bibinfo {pages}
  {053601} (\bibinfo {year} {2014})}\BibitemShut {NoStop}%
\bibitem [{\citenamefont {Baur}\ \emph {et~al.}(2014)\citenamefont {Baur},
  \citenamefont {Tiarks}, \citenamefont {Rempe},\ and\ \citenamefont
  {D\"urr}}]{baur2014sing}%
  \BibitemOpen
  \bibfield  {author} {\bibinfo {author} {\bibfnamefont {S.}~\bibnamefont
  {Baur}}, \bibinfo {author} {\bibfnamefont {D.}~\bibnamefont {Tiarks}},
  \bibinfo {author} {\bibfnamefont {G.}~\bibnamefont {Rempe}}, \ and\ \bibinfo
  {author} {\bibfnamefont {S.}~\bibnamefont {D\"urr}},\ }\href {\doibase
  10.1103/PhysRevLett.112.073901} {\bibfield  {journal} {\bibinfo  {journal}
  {Phys. Rev. Lett.}\ }\textbf {\bibinfo {volume} {112}},\ \bibinfo {pages}
  {073901} (\bibinfo {year} {2014})}\BibitemShut {NoStop}%
\bibitem [{\citenamefont {Tiarks}\ \emph {et~al.}(2014)\citenamefont {Tiarks},
  \citenamefont {Baur}, \citenamefont {Schneider}, \citenamefont {D\"urr},\
  and\ \citenamefont {Rempe}}]{tiar2014sing}%
  \BibitemOpen
  \bibfield  {author} {\bibinfo {author} {\bibfnamefont {D.}~\bibnamefont
  {Tiarks}}, \bibinfo {author} {\bibfnamefont {S.}~\bibnamefont {Baur}},
  \bibinfo {author} {\bibfnamefont {K.}~\bibnamefont {Schneider}}, \bibinfo
  {author} {\bibfnamefont {S.}~\bibnamefont {D\"urr}}, \ and\ \bibinfo {author}
  {\bibfnamefont {G.}~\bibnamefont {Rempe}},\ }\href {\doibase
  10.1103/PhysRevLett.113.053602} {\bibfield  {journal} {\bibinfo  {journal}
  {Phys. Rev. Lett.}\ }\textbf {\bibinfo {volume} {113}},\ \bibinfo {pages}
  {053602} (\bibinfo {year} {2014})}\BibitemShut {NoStop}%
\bibitem [{\citenamefont {Siegman}(1986)}]{sieg1986lase}%
  \BibitemOpen
  \bibfield  {author} {\bibinfo {author} {\bibfnamefont {A.~E.}\ \bibnamefont
  {Siegman}},\ }\href@noop {} {\emph {\bibinfo {title} {Lasers}}}\ (\bibinfo
  {publisher} {University Science Books},\ \bibinfo {address} {Sausalito},\
  \bibinfo {year} {1986})\BibitemShut {NoStop}%
\bibitem [{\citenamefont {Berry}(1987)}]{berr1987adia}%
  \BibitemOpen
  \bibfield  {author} {\bibinfo {author} {\bibfnamefont {M.~V.}\ \bibnamefont
  {Berry}},\ }\href {\doibase 10.1080/09500348714551321} {\bibfield  {journal}
  {\bibinfo  {journal} {J. Mod. Opt}\ }\textbf {\bibinfo {volume} {34}},\
  \bibinfo {pages} {1401} (\bibinfo {year} {1987})}\BibitemShut {NoStop}%
\bibitem [{\citenamefont {Amthor}\ \emph {et~al.}(2007)\citenamefont {Amthor},
  \citenamefont {Reetz-Lamour}, \citenamefont {Giese},\ and\ \citenamefont
  {Weidem\"uller}}]{amth2007mode}%
  \BibitemOpen
  \bibfield  {author} {\bibinfo {author} {\bibfnamefont {T.}~\bibnamefont
  {Amthor}}, \bibinfo {author} {\bibfnamefont {M.}~\bibnamefont
  {Reetz-Lamour}}, \bibinfo {author} {\bibfnamefont {C.}~\bibnamefont {Giese}},
  \ and\ \bibinfo {author} {\bibfnamefont {M.}~\bibnamefont {Weidem\"uller}},\
  }\href {\doibase 10.1103/PhysRevA.76.054702} {\bibfield  {journal} {\bibinfo
  {journal} {Phys. Rev. A}\ }\textbf {\bibinfo {volume} {76}},\ \bibinfo
  {pages} {054702} (\bibinfo {year} {2007})}\BibitemShut {NoStop}%
\bibitem [{\citenamefont {Beterov}\ \emph {et~al.}(2009)\citenamefont
  {Beterov}, \citenamefont {Ryabtsev}, \citenamefont {Tretyakov},\ and\
  \citenamefont {Entin}}]{bete2009quas}%
  \BibitemOpen
  \bibfield  {author} {\bibinfo {author} {\bibfnamefont {I.~I.}\ \bibnamefont
  {Beterov}}, \bibinfo {author} {\bibfnamefont {I.~I.}\ \bibnamefont
  {Ryabtsev}}, \bibinfo {author} {\bibfnamefont {D.~B.}\ \bibnamefont
  {Tretyakov}}, \ and\ \bibinfo {author} {\bibfnamefont {V.~M.}\ \bibnamefont
  {Entin}},\ }\href {\doibase 10.1103/PhysRevA.79.052504} {\bibfield  {journal}
  {\bibinfo  {journal} {Phys. Rev. A}\ }\textbf {\bibinfo {volume} {79}},\
  \bibinfo {pages} {052504} (\bibinfo {year} {2009})}\BibitemShut {NoStop}%
\bibitem [{\citenamefont {{Cohen-Tannoudji, Claude}}\ \emph
  {et~al.}(2004)\citenamefont {{Cohen-Tannoudji, Claude}}, \citenamefont
  {{Dupont-Roc, Jacques}},\ and\ \citenamefont {{Grynberg,
  Gilbert}}}]{cohe2004atom}%
  \BibitemOpen
  \bibfield  {author} {\bibinfo {author} {\bibnamefont {{Cohen-Tannoudji,
  Claude}}}, \bibinfo {author} {\bibnamefont {{Dupont-Roc, Jacques}}}, \ and\
  \bibinfo {author} {\bibnamefont {{Grynberg, Gilbert}}},\ }\href@noop {}
  {\emph {\bibinfo {title} {Atom - {Photon} {Interactions}: {Basic} {Processes}
  and {Applications}}}}\ (\bibinfo  {publisher} {Wiley},\ \bibinfo {year}
  {2004})\BibitemShut {NoStop}%
\bibitem [{\citenamefont {Brecha}\ \emph {et~al.}(1999)\citenamefont {Brecha},
  \citenamefont {Rice},\ and\ \citenamefont {Xiao}}]{brec1999n}%
  \BibitemOpen
  \bibfield  {author} {\bibinfo {author} {\bibfnamefont {R.~J.}\ \bibnamefont
  {Brecha}}, \bibinfo {author} {\bibfnamefont {P.~R.}\ \bibnamefont {Rice}}, \
  and\ \bibinfo {author} {\bibfnamefont {M.}~\bibnamefont {Xiao}},\ }\href@noop
  {} {\bibfield  {journal} {\bibinfo  {journal} {Phys. Rev. A}\ }\textbf
  {\bibinfo {volume} {59}},\ \bibinfo {pages} {2392} (\bibinfo {year}
  {1999})}\BibitemShut {NoStop}%
\end{thebibliography}

\end{document}